\documentclass[reprint,showpacs,preprintnumbers,amsmath,amssymb,aps,pra,nofootinbib,superscriptaddress,floatfix]{revtex4-1}
\usepackage{bbold}
\usepackage{hyperref}

\usepackage{graphicx}

\usepackage{color}
\def\comment#1{{#1}}
\def\highlight#1{{#1}}
\def\Highlight#1{{#1}}
\newcommand{\appsection}[1]{\section{\uppercase{#1}}}

\begin{document}

\title{Overcoming loss of contrast in atom interferometry due to gravity gradients}

\author{Albert Roura}
\affiliation{Institut f\"ur Quantenphysik, Universit\"at Ulm, Albert-Einstein-Allee 11,
89081 Ulm, Germany}

\author{Wolfgang Zeller}
\affiliation{Institut f\"ur Quantenphysik, Universit\"at Ulm, Albert-Einstein-Allee 11,
89081 Ulm, Germany}

\author{Wolfgang P.\ Schleich}
\affiliation{Institut f\"ur Quantenphysik, Universit\"at Ulm, Albert-Einstein-Allee 11,
89081 Ulm, Germany}
\affiliation{Institute for Quantum Science and Engineering (IQSE), and Department
of Physics and Astronomy, Texas A\&M University, College Station,
Texas 77843-4242, USA}

\date{\today}
\begin{abstract}
Long-time atom interferometry is instrumental to various high-precision measurements of fundamental physical properties, including tests of the equivalence principle. Due to rotations and gravity gradients, the classical trajectories characterizing the motion of the wave packets for the two branches of the interferometer do not close in phase space, an effect which increases significantly with the interferometer time.
The relative displacement between the interfering wave packets in such open interferometers leads to a fringe pattern in the density profile at each exit port and a loss of contrast in the oscillations of the integrated particle number as a function of the phase shift. Paying particular attention to gravity gradients, we present a simple mitigation strategy involving small changes in the timing of the laser pulses which is very easy to implement.
A useful representation-free description of the state evolution in an atom interferometer is introduced and employed to analyze the loss of contrast and mitigation strategy in the general case. (As a by-product, a remarkably compact derivation of the phase-shift in a general light-pulse atom interferometer is provided.)
Furthermore, exact results are obtained for (pure and mixed) Gaussian states which allow a simple interpretation in terms of the alignment of Wigner functions in phase-space. Analytical results are also obtained for expanding Bose-Einstein condensates within the time-dependent Thomas-Fermi approximation. Finally, a combined strategy for rotations and nonaligned gravity gradients is considered as well.
\end{abstract}


\maketitle

\section{Introduction}
\label{sec:introduction}

In this article we analyze in detail the loss of integrated contrast in open atom interferometers, where classical phase-space trajectories do not close, with particular emphasis on the effects of gravity gradients. A simple but effective mitigation strategy to overcome such loss of contrast is presented.

The potential of light-pulse atom interferometry \cite{borde89,kasevich91} for high-precision measurements has been amply demonstrated with its successful implementation in extremely sensitive inertial sensors, including gyroscopes \cite{gustavson00,canuel06,tackmann12}, gradiometers \cite{mcguirk02} and the currently most precise absolute gravimeters \cite{peters99,peters01,merlet10}.
It has already found applications in accurate measurements of fundamental constants \cite{fixler07,lamporesi09,rosi14,weiss93,wicht02,cadoret08} and tests of fundamental properties \cite{mueller08c,bouchendira11,bouchendira13}, and it is a key ingredient in plans for future tests of the equivalence principle in space \cite{aguilera13,ste-quest2}
\highlight{(atom-interferometry-based experiments have already been performed on the ground \cite{fray04,bonnin13,schlippert14,tarallo14}),}
next-generation satellite geodesy missions \cite{carraz13}
or even alternative proposals for gravitational-wave detection \cite{hogan11}.
Achieving higher sensitivities requires extended interferometer times (the phase shift generated by accelerations, for instance, grows with the square of the interferometer time). Remarkable progress in this direction has recently been made with atomic fountains on the ground \cite{dickerson13}. Nevertheless, the natural environment for reaching even longer interferometer times are the microgravity conditions provided by space missions \cite{cal,aguilera13} and being preliminarily investigated in parabolic flights \cite{geiger11}, drop towers \cite{vanzoest10,muentinga13} and sounding rockets \cite{maius}.

Measurements based on atom interferometry like those mentioned above typically involve keeping track of the oscillations in the total number of atoms at each exit port as certain parameters, such as the half-interferometer time $T$, are changed. Besides other challenges of practical nature, the long interferometer times planned for future space missions face a more fundamental limitation due to gravity gradients. These cause a relative shift in position and momentum between the pair of wave packets contributing to the superposition at each exit port, which leads to a reduction of the integrated contrast (i.e.\ a smaller amplitude of the oscillations in the integrated atom number at each port) and a corresponding decrease in sensitivity.
As an example, plans for the European Space Agency's STE-QUEST mission \cite{aguilera13} have estimated a 40\% loss of contrast due to gravity gradients despite a required effective temperature for the atom cloud below $70\, \text{pK}$, and this is the main driver behind the need for such a narrow momentum distribution (although it has other added benefits on diffraction efficiencies or systematics associated with the Rabi frequency).

Such a loss of contrast is, in fact, a generic feature of open interferometers, where the classical trajectories characterizing the motion of the wave packets do not close in phase space and the resulting relative displacements in position and momentum, $\delta\boldsymbol{\mathcal{R}}$ and $\delta\boldsymbol{\mathcal{P}}$, cause a reduction of the quantum overlap between the interfering states in each exit port. In position representation this reduction is due to two effects: first, a decreasing spatial overlap of the two envelopes; second, spatial oscillations in the product of one wave function and the other's complex conjugate, whose integral over space gives the quantum overlap of the two states. As we will see in forthcoming sections, in situations relevant for atom interferometry (at least before implementing our proposed mitigation strategy) the second effect typically becomes important well before the spatial displacement becomes comparable to the size of the envelopes and the first effect starts to play a role.

The spatial oscillations described in the previous paragraph give rise to a fringe pattern in the spatial density profile of each exit port which is closely related to the contrast loss. This suggests devising a simple mitigation strategy that consists in eliminating the fringe pattern by means of a small but properly chosen change $\delta T$ in the timing of the last beam-splitter pulse (i.e.\ the time at which it is applied). Doing so hardly alters the the momentum displacement $\delta\boldsymbol{\mathcal{P}}$, but it can be used to change $\delta\boldsymbol{\mathcal{R}}$ in such a way that its effect essentially cancels that due to $\delta\boldsymbol{\mathcal{P}}$ and the fringe spacing becomes much larger than the size of the envelope.
In this paper we will mainly focus on gravity gradients aligned with the direction of the laser beams, but will express our results using a vector notation that makes them directly applicable to the general case (when the principal axes of the gravity gradient tensor are no longer aligned with the lasers) as well. However, since nonaligned gravity gradients produce also displacements along the transverse directions, adjusting the timing of the last beam-splitter, which only generates displacements along the longitudinal direction, is not enough. As explained in Sec.~\ref{sec:nonaligned}, in that case one can deal with the transverse displacements by using the same kind of scheme based on a tip-tilt mirror which has already been employed by several groups to compensate the effects of rotations \cite{hogan08,lan12,hauth12,dickerson13,sugarbaker13}. 

Even though in many of the calculations that will be presented in this article an exact treatment of the gravity gradients is possible (at least for time-independent ones), the results are simpler and more transparent when treating the gravity gradient tensor $\Gamma$ perturbatively. This is, in fact, an excellent approximation for the calculation of the contrast because the parameter $\| \Gamma \|\, T^2$ controlling the perturbative expansion (where $\| \Gamma \|$ denotes the largest eigenvalue in absolute value) is typically much smaller than unity: even for half-interferometer times of $T=10\, \text{s}$ it is still of the order of $10^{-4}$. Therefore, one can perform a perturbative calculation of the classical trajectories that characterize the motion of the wave packets to determine the relative displacement between the corresponding pair of wave packets at each exit port, which is of order $\Gamma \, T^2$ (although somewhat enhanced by a factor $\mathbf{v}_\text{rec} \, T$).
Furthermore, when determining how the shape of the wave packets evolves in time (or, more precisely, how the centered wave packets defined in Appendix~\ref{sec:rep_free} evolve), it is sufficient to consider their free evolution (neglecting the gravity gradients). This is because in the absence of relative displacement the states of the two wave packets coincide up to a phase and have maximum quantum overlap (hence, independent of $\Gamma$). Therefore, since the relative displacement is itself of order $\Gamma$, the effect of the gravity gradients on the evolution of the centered wave packet would generate corrections to the quantum overlap of higher order in $\Gamma$.

Although the main goal of this article is to study the loss of integrated contrast in open interferometers and how to overcome it, it is worth pointing out that many of the techniques and results presented here can also be directly used to explain how gravity gradients, rotations and changes in pulse timing modify the density profile observed at the exit ports of an atom interferometer \cite{zeller14a}, as measured in the experiments reported in Refs.~\cite{muentinga13,dickerson13,sugarbaker13}. Moreover, as a side result we provide in Appendix~\ref{sec:rep_free} a remarkably compact derivation of the general formula for the phase shift in presence of time-dependent gravity gradients and uniform force fields.

The paper is organized as follows.
The general framework, the reasons for the loss of contrast in open interferometers and the mitigation strategy are presented in Sec.~\ref{sec:general}, which contains our central findings. A detailed analysis for pure Gaussian states, with exact results and quantitative examples, is provided in Sec.~\ref{sec:gaussian}. They illustrate the phase-space interpretation of the loss of contrast and the mitigation strategy in terms of the alignment of squeezed Wigner functions. In addition, in Sec.~\ref{sec:gaussian_mixed} we discuss the new features connected with mixed states in general and their particularization to the Gaussian case. Interferometers based on expanding Bose-Einstein condensates (BECs) are investigated in Sec.~\ref{sec:BEC}, where we give analytical results using the scaling approach and the time-dependent Thomas-Fermi approximation as well as quantitative examples. The case of nonaligned gravity gradients is considered in Sec.~\ref{sec:nonaligned}. Finally, our results and future perspectives are discussed in Sec.~\ref{sec:conclusions}.
A~number of technical aspects have been included in the appendices. The classical phase-space trajectories (and the relative displacements) in the presence of gravity gradients, uniform force fields and taking into account the laser kicks are computed in Appendix~\ref{sec:trajectories}.
A representation-free description for quadratic Hamiltonians of the state evolution in the atom interferometer is presented in Appendix~\ref{sec:rep_free}, including a very compact derivation of the general result for the phase shift. In Appendix~\ref{sec:late-time} we obtain the late-time result for the free evolution of a wave packet.
Finally, the Gross-Pitaevskii equation and its solution for expanding BECs within the framework of the scaling approach and the time-dependent Thomas-Fermi approximation are briefly reviewed in Appendix~\ref{sec:BEC_evolution} and its connection with free evolution at late times is discussed in some detail.


%
%
%
%
%
%
%
%
%

\section{Loss of contrast and mitigation strategy}
\label{sec:general}

\subsection{Representation-free description for atom interferometers}
\label{sec:representation-free}

Throughout the paper we will focus on light-pulse atom interferometers, based on the diffraction of matter waves by time-modulated laser pulses which induce transitions with a fixed momentum transfer (and possibly changing the internal state). By selecting their duration and intensity, one can have $\pi$ and $\pi/2$ pulses acting respectively as mirrors and beam splitters. For simplicity we will assume idealized beam splitters and mirrors with 100\% efficiency, no dispersive effects within the relevant momentum range for the atoms (which would otherwise change the shape of the wave packet) and sufficiently short pulse duration so that the evolution of the external degrees of freedom can be neglected during the pulses. Furthermore, we will assume a perfect read-out of the two exit ports thanks to good spatial separation or state labeling. Finally, when considering quantum degenerate gases, their density during the interferometer sequence should be low enough so that nonlinear interactions can be neglected and their dynamics can be accurately described in terms of the Schr\"odinger equation for noninteracting particles.

Besides the interaction with the laser fields during the short duration of the laser pulses, the dynamics of the atoms will be governed by the following quadratic Hamiltonian:
\begin{equation}
\hat{H} = \frac{1}{2m} \hat{\mathbf{p}}^\text{T} \hat{\mathbf{p}}
- \frac{m}{2} \hat{\mathbf{x}}^\text{T} \Gamma (t)\, \hat{\mathbf{x}}
- m\, \mathbf{g}^\text{T} (t) \, \hat{\mathbf{x}} + V_0(t)
\label{eq:hamiltonian}.
\end{equation}
It can describe the effect of time-dependent gravity gradients and uniform force (or acceleration) fields, characterized respectively by the tensor $\Gamma (t)$ and the vector $\mathbf{g} (t)$,
\highlight{as well as different internal energies, characterized by the spatially independent potential $V_0(t)$}. It can also be employed to describe the effect of rotations from the point of view of the non-rotating frame: one needs then to take into account the rotation of the momentum transfers $\hbar \mathbf{k}_{i}$ associated with the different laser pulses, and possibly of $\Gamma (t)$ and $\mathbf{g}(t)$ too \cite{kleinert14a}.

As shown in Appendix~\ref{sec:rep_free}, for quadratic Hamiltonians like that of Eq.~\eqref{eq:hamiltonian} there is a useful representation-free description for the time evolution of the atoms within the interferometer in terms of the evolution of a \emph{centered} wave packet plus classical trajectories which include the kicks from the laser pulses and characterize the motion of each wave packet contributing to the superposition.
\highlight{(A similar construction for the free propagation between laser pulses has been considered in Refs.~\cite{borde92,hogan08,dimopoulos08}).}
For interferometers consisting of an initial and a final beam splitter together with several $\pi$ pulses in between (the Mach-Zehnder configuration being a typical example) the state of the system during the interferometer phase (between the first and last beam splitters) is given by
\begin{equation}
|\psi (t)\rangle =
\frac{1}{\sqrt{2}} \Big[ e^{i \Phi_a} \hat{\mathcal{D}}(\boldsymbol{\chi}_a)  |\psi_\text{c} (t) \rangle
+ e^{i \Phi_b} \hat{\mathcal{D}}(\boldsymbol{\chi}_b)  |\psi_\text{c} (t) \rangle \Big]
\label{1},
\end{equation}
where $|\psi_\text{c} (t) \rangle$ is a centered wave packet evolving in a purely quadratic potential and we have omitted the time dependence of the phases and displacement vectors to avoid an unnecessarily cumbersome notation. The displacement operator $\hat{\mathcal{D}}(\boldsymbol{\chi})$ is defined as
\begin{equation}
\hat{\mathcal{D}}(\boldsymbol{\chi}) = e^{-\frac{i}{\hbar} \boldsymbol{\chi}^\text{T} J \, \hat{\boldsymbol{\xi}}}
\label{eq:displacement} ,
\end{equation}
where we used a vector notation for phase-space quantities, so that $\hat{\boldsymbol{\xi}} = (\hat{\mathbf{x}},\hat{\mathbf{p}})^\text{T}$, and introduced the symplectic form
\begin{equation}
J = \left( \begin{array}{cc} 0 & \mathbb{1} \\ -\mathbb{1} & 0 \end{array} \right)
\label{3}.
\end{equation}
The displacement vectors
$\boldsymbol{\chi} (t) = \big(\boldsymbol{\mathcal{R}}(t), \boldsymbol{\mathcal{P}}(t) \big)^\text{T}$ in Eq.~\eqref{1} correspond to the phase-space vectors for the classical trajectories associated with each branch, $a$ and $b$, of the interferometer.

\begin{figure}[h]
\begin{center}
\includegraphics[width=8cm]{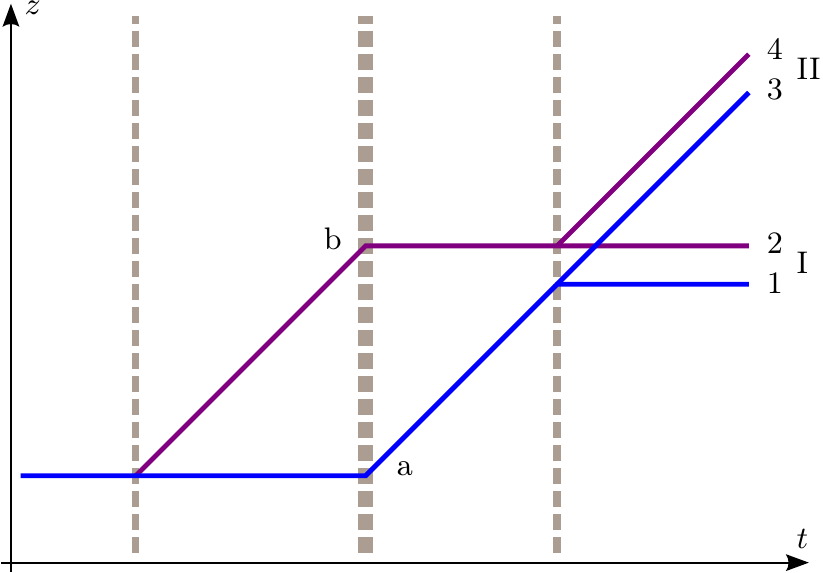}
\end{center}
\caption{Classical trajectories characterizing the motion of the wave packets in a Mach-Zehnder interferometer. The two branches of the interferometer are labeled as $a$ and $b$, whereas the trajectories of the two interfering wave packets at each exit port are labeled as $\{1,2\}$ and $\{3,4\}$ respectively.}
\label{fig:interferometer1}
\end{figure}

Provided that the two exit ports can be perfectly distinguished, due to absence of spatial overlap or different internal states, only two wave packets contribute to each port, as depicted in Fig.~\ref{fig:interferometer1}. Therefore, we have the following superposition at one of the exit ports (after the second beam splitter):
\begin{align}
|\psi_\text{I} (t)\rangle &=
\frac{1}{2} \Big[ e^{i \Phi_1} \hat{\mathcal{D}}(\boldsymbol{\chi}_1)  |\psi_\text{c} (t) \rangle
+ e^{i \Phi_2} \hat{\mathcal{D}}(\boldsymbol{\chi}_2)  |\psi_\text{c} (t) \rangle \Big] \nonumber \\
&= \frac{1}{2} e^{i \Phi_1} \hat{\mathcal{D}}(\boldsymbol{\chi}_1)
\Big[ 1 + e^{i \delta\Phi} \hat{\mathcal{D}}(\delta\boldsymbol{\chi}) \Big] |\psi_\text{c} (t) \rangle
\label{4},
\end{align}
with $\delta\boldsymbol{\chi} = \boldsymbol{\chi}_2 - \boldsymbol{\chi}_1$ and $\delta\Phi = \Phi_2 - \Phi_1 + \boldsymbol{\chi}_1^\text{T} J \, \boldsymbol{\chi}_2 / 2 \hbar$, where the extra phase arises from the composition of displacement operators:
\begin{equation}
\hat{\mathcal{D}}(\boldsymbol{-\chi}_1) \, \hat{\mathcal{D}}(\boldsymbol{\chi}_2)
= e^{\frac{i}{2 \hbar} \boldsymbol{\chi}_1^\text{T} J \, \boldsymbol{\chi}_2} \,
\hat{\mathcal{D}}(\delta\boldsymbol{\chi})
\label{5}.
\end{equation}
The result is analogous for the state $|\psi_\text{II} (t)\rangle$ at the second exit port.

While we will focus on pure states throughout this section, mixed sates can be treated as an incoherent ensemble of pure states, as explained in detail in Sec.~\ref{sec:gaussian_mixed}.

\subsection{Loss of contrast in open interferometers}
\label{sec:contrast_loss}

\begin{figure}[h]
\begin{center}
\includegraphics[width=8cm]{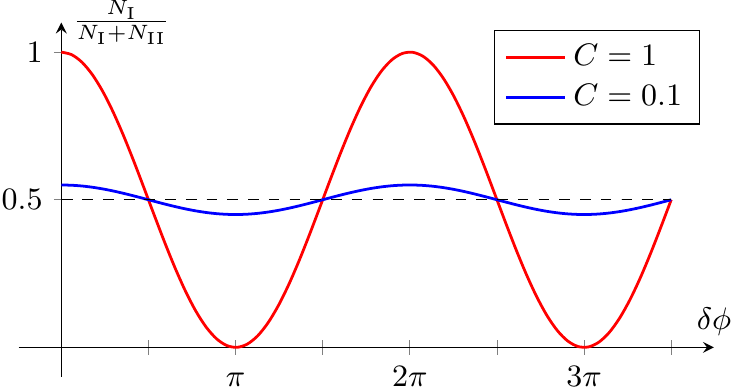}
\end{center}
\caption{Oscillations of the fraction of atoms in exit port I as a function of the phase shift $\delta\phi$. The amplitude of the oscillations is maximal (red curve) for $C=1$ and smaller (blue curve) for $C < 1$.}
\label{fig:contrast}
\end{figure}

We will employ the term ``open interferometer'' whenever $\delta\boldsymbol{\chi} \neq \mathbf{0}$ at the time of detection. This will lead in general to a loss of contrast when considering the integrated number of particles at each exit port as a function of the phase shift between the interferometer branches.
Indeed, the fraction of atoms in the first exit port is given by $\big\langle \psi_\text{I} (t) \big| \psi_\text{I} (t) \big\rangle$ and from Eq.~\eqref{4} one can easily obtain
\begin{equation}
\frac{N_\text{I}}{N_\text{I} + N_\text{II}}
= \big\langle \psi_\text{I} (t) \big| \psi_\text{I} (t) \big\rangle
= \frac{1}{2} \big(1 + C \cos \delta\phi \big)
\label{6},
\end{equation}
with
\begin{equation}
C = \Big| \big\langle \psi_\text{c} (t) \big| \hat{\mathcal{D}}(\delta\boldsymbol{\chi})
\big| \psi_\text{c} (t) \big\rangle \Big| \leq 1
\label{7}.
\end{equation}
Therefore, $\delta\boldsymbol{\chi} = 0$ implies $C=1$ and full oscillations, between 0 and $N$, in the total atom number at each exit port. In contrast, $\delta\boldsymbol{\chi} \neq 0$ causes a contrast reduction%
\footnote{\Highlight{This can be interpreted as a consequence of the (partial) distinguishability of the two states superimposed at the exit port. Entanglement to an additional quantum system or degrees of freedom can also  lead to the distinguishability of the two states, with analogous consequences \cite{englert96}. In the latter case, however, neither the strategy proposed in Sec.~\ref{sec:strategy} nor the measurement of the fringe pattern in the density profile as discussed in Sec.~\ref{sec:conclusions} can be employed to recover the lost quantum coherence and to determine the phase shift between the two interferometer branches.}}
with smaller oscillations around $N/2$ of the atom number at each port, as illustrated in Fig.~\ref{fig:contrast}.
(Completely analogous results hold for the second exit port.)
Note that in general the expectation value $\big\langle \psi_\text{c} (t) \big| \hat{\mathcal{D}}(\delta\boldsymbol{\chi}) \big| \psi_\text{c} (t) \big\rangle$ is complex and its phase gives a contribution to the phase $\delta\phi$ in Eq.~\eqref{6} in addition to the phase shift $\delta\Phi$ in Eq.~\eqref{4}.

In order to gain further insight on the loss of contrast in open interferometers and help us devise a suitable mitigation strategy, it is instructive to consider the situation in position representation. Choosing the origin of coordinates so that $\boldsymbol{\mathcal{R}}_1 = 0$ to ease the notation, the probability density associated with the state in Eq.~\eqref{4} becomes
\begin{align}
\big|\psi_\text{I} (\mathbf{x},t) \big|^2
=& \frac{1}{4} \Big( \big|\psi_\text{c} (\mathbf{x},t) \big|^2
+ \big|\psi_\text{c} (\mathbf{x}-\delta\boldsymbol{\mathcal{R}},t) \big|^2 \Big) \nonumber \\
& + \frac{1}{2} \mathrm{Re} \Big( e^{i \, \delta\Phi'}
e^{\frac{i}{\hbar} \delta \boldsymbol{\mathcal{P}}^\text{T} \mathbf{x} } \,
\psi_\text{c}^* (\mathbf{x},t) \,
\psi_\text{c} (\mathbf{x}-\delta\boldsymbol{\mathcal{R}},t) \Big)
\label{eq:prob_density1} ,
\end{align}
where some spatially independent phase terms (i.e.\ independent of $\mathbf{x}$) have been absorbed in the phase $\delta\Phi'$ to make the expression more compact. It is particularly illuminating to consider the result for free evolution at sufficiently late times, which is analyzed in Appendix~\ref{sec:late-time} and is relevant for typical long-time interferometry applications. Using the state in Eq.~\eqref{C3}, Eq.~\eqref{eq:prob_density1} becomes
\begin{equation}
\big|\psi_\text{I} (\mathbf{x},t) \big|^2 \approx \frac{1}{4}
\Big| 1 + e^{i \delta\Phi''} e^{\frac{i}{\hbar} \left( \delta \boldsymbol{\mathcal{P}}
- \frac{m}{\Delta t} \delta \boldsymbol{\mathcal{R}} \right)^\text{T} \mathbf{x}} \Big|^2
\big|\psi_\text{c} (\mathbf{x},t) \big|^2
\label{eq:prob_density2} ,
\end{equation}
where some spatially independent phase terms have again been absorbed in the phase $\delta\Phi''$ and  \comment{$\Delta t = t-t_0$}. For simplicity we have considered here the case in which there is no spatially dependent contribution to the phase from the factor $\tilde{\psi} (m\, \mathbf{x}/\Delta t,t_0)$ in Eq.~\eqref{C3}. Moreover, we have neglected the shift by $\delta \boldsymbol{\mathcal{R}}$ of the envelope, which is a reasonable approximation whenever the envelope $\big|\psi_\text{c} (\mathbf{x},t) \big|$ does not vary too rapidly in space and its size is much larger than $\delta \boldsymbol{\mathcal{R}}$. (This approximation, which is often an excellent one in typical situations of interest, will be discussed in detail in Secs.~\ref{sec:gaussian_examples} and \ref{sec:BEC_examples}.)
Eq.~\eqref{eq:prob_density2} reveals the existence of a fringe pattern in the density profile of each exit port with a fringe spacing $\lambda_\text{fr}$ characterized by
\begin{equation}
\frac{2\pi}{\lambda_\text{fr}} \boldsymbol{\hat{\mathbf{n}}}
= \frac{1}{\hbar} \left( \delta \boldsymbol{\mathcal{P}}
- \frac{m}{\Delta t} \delta \boldsymbol{\mathcal{R}} \right)
\label{9}.
\end{equation}

\begin{figure}[h]
\begin{center}
\includegraphics[width=4cm]{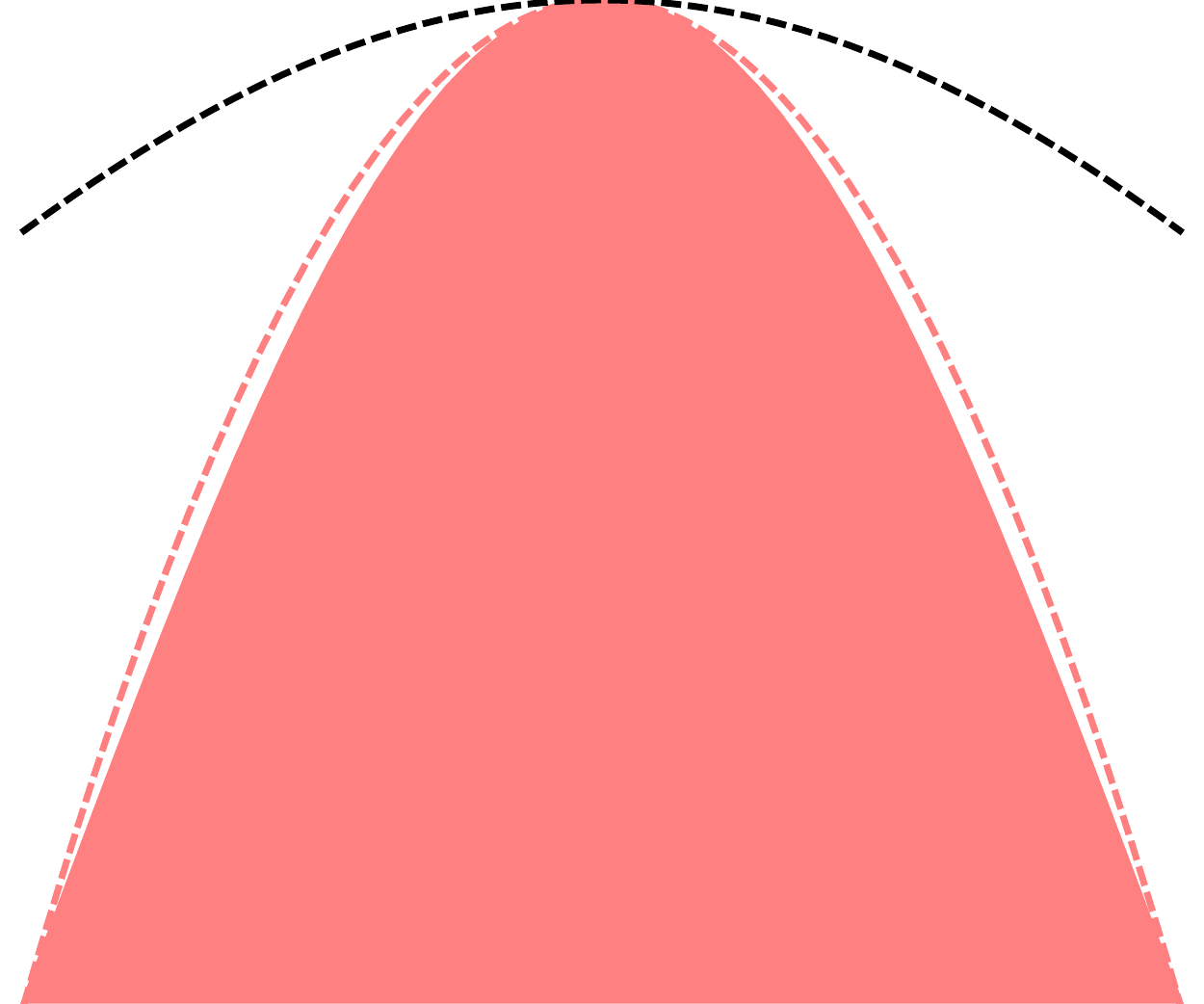}
\includegraphics[width=4cm]{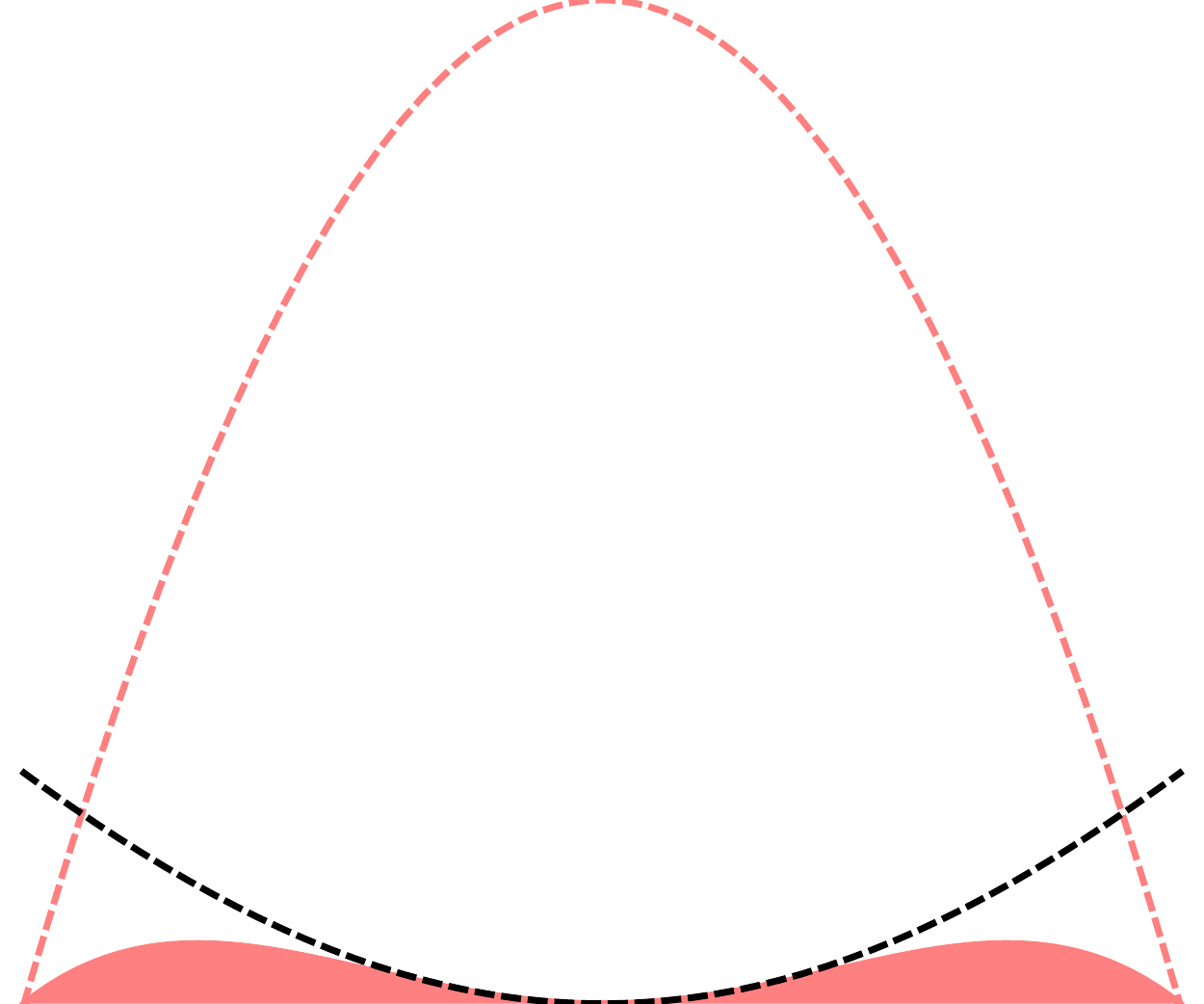} \\
\vspace{0.5cm}
\includegraphics[width=4cm]{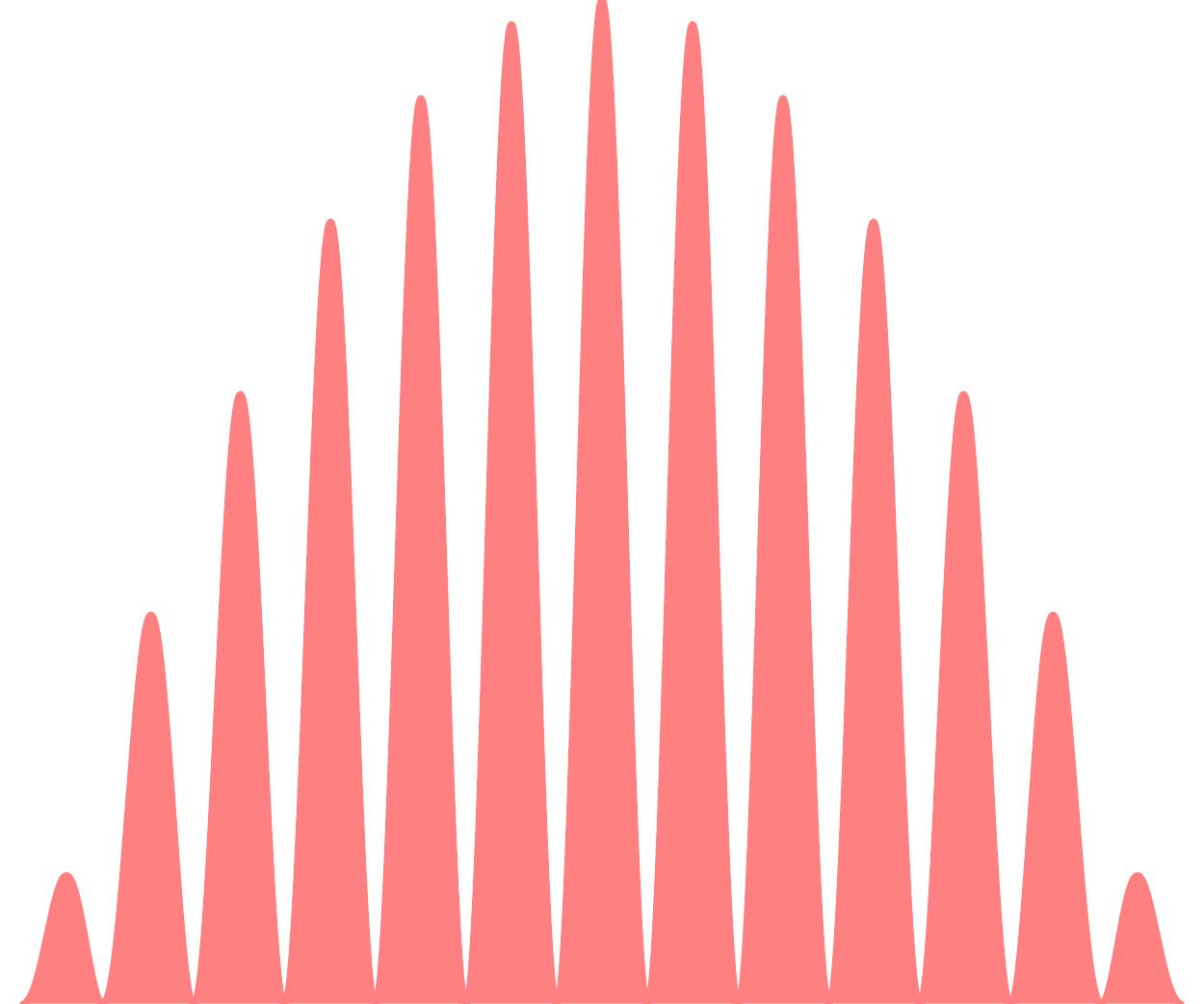}
\includegraphics[width=4cm]{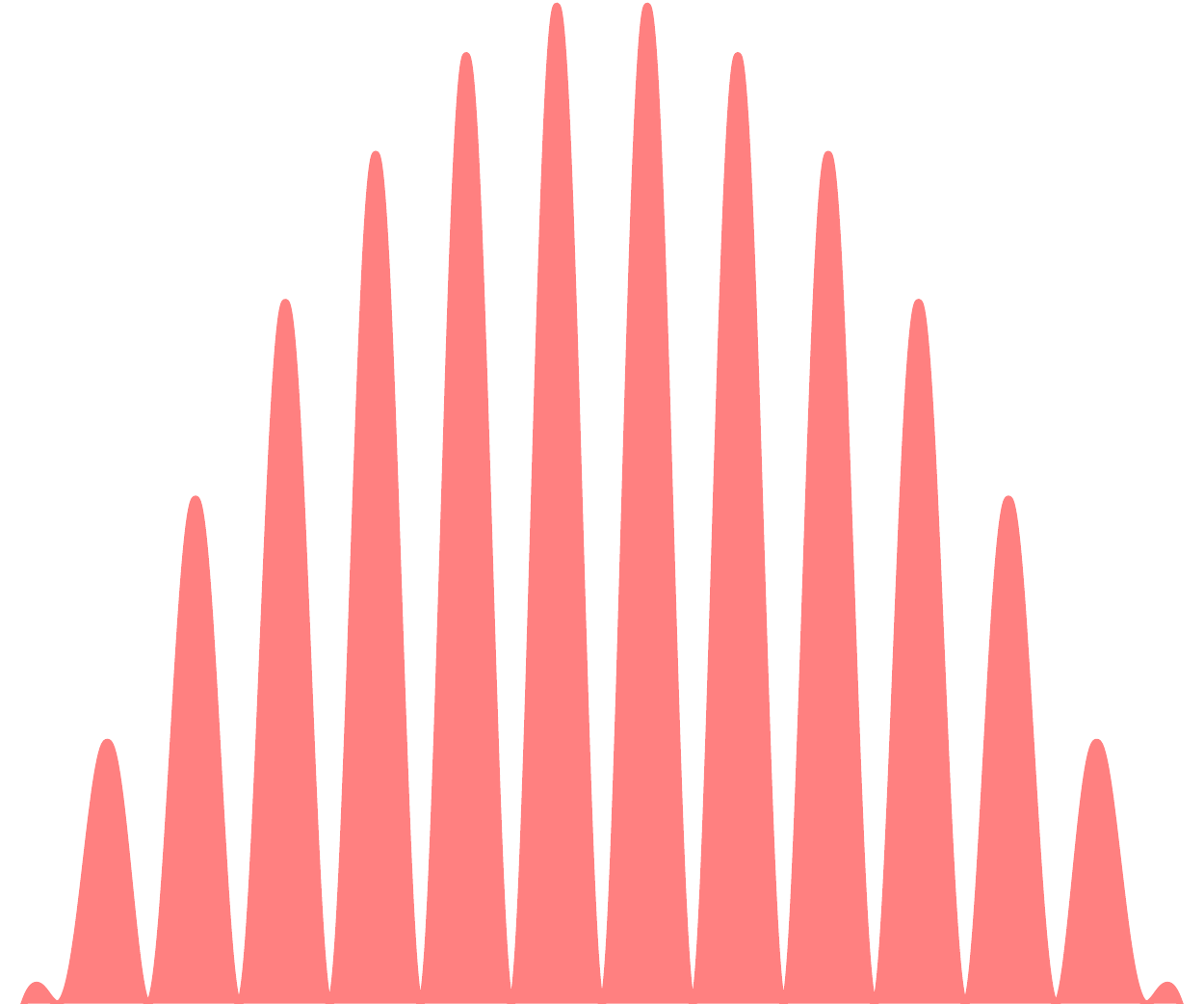}
\end{center}
\caption{The two upper pictures depict the density profile at the exit port for $\delta \phi = 0$ and $\delta \phi = \pi$ when the size of the fringes (dashed dark line) is much larger than the envelope (dashed light line); the port changes from almost completely ``bright'' to almost completely ``dark'' as far as the integrated particle number is concerned. The lower ones are analogous pictures when the fringe spacing is much smaller than the size of the envelope; the integrated particle number at the exit port changes very little as $\delta \phi$ varies.} 
\label{fig:fringes}
\end{figure}

Integrating Eq.~\eqref{eq:prob_density1} over space, one gets the analog of Eqs.~\eqref{6}-\eqref{7} in position representation with
\begin{equation}
C = \left| \int d^3 x \, e^{\frac{i}{\hbar} \delta \boldsymbol{\mathcal{P}}^\text{T} \mathbf{x}}
\, \psi_\text{c}^* (\mathbf{x}+\delta\boldsymbol{\mathcal{R}}/2,t) \,
\psi_\text{c} (\mathbf{x}-\delta\boldsymbol{\mathcal{R}}/2,t) \right| 
\label{10}.
\end{equation}
where we have shifted the integration variable $\mathbf{x}$ by $\delta\boldsymbol{\mathcal{R}}/2$ to write the result in a more convenient form.
Specializing to the state in Eq.~\eqref{C3} and making use of the same assumptions employed when deriving Eq.~\eqref{eq:prob_density2}, which includes neglecting the shift of the envelopes, Eq.~\eqref{10} becomes%
\footnote{The relative displacement between the envelopes as well as the possibility of having spatially dependent contributions to the phase from $\tilde{\psi} (m\, \mathbf{x}/\Delta t,t_0)$ and the corresponding corrections to the fringe spacing~\eqref{9}
will be taken into account when studying the evolution of Gaussian states and of expanding BECs in Secs.~\ref{sec:gaussian} and \ref{sec:BEC} respectively.}
\begin{equation}
C \approx \left| \int d^3 x \exp\left[ \frac{i}{\hbar} \left( \delta \boldsymbol{\mathcal{P}}
- \frac{m}{\Delta t} \delta \boldsymbol{\mathcal{R}} \right)^\text{T} \! \mathbf{x} \right] 
\big|\psi_\text{c} (\mathbf{x},t) \big|^2  \right|
\label{11}.
\end{equation}
It is then easy to see that when the fringe spacing is much larger than the size of the envelope, as shown in the upper row of Fig.~\ref{fig:fringes}, the port goes from almost completely bright to almost completely dark as $\delta\phi$ changes, and the integrated particle number at each port oscillates from $0$ to $N$ (with $C \approx 1$).
On the other hand, for fringe spacing much smaller than the size of the envelope, as shown in the lower row of Fig.~\ref{fig:fringes}, $C \ll 1$ and the integrated particle number at each port exhibits small amplitude oscillations around $N/2$.
Besides providing an intuitive understanding for the loss of contrast in open interferometers, having identified the existence of a fringe pattern in the density profile at each exit port as the culprit can serve as a useful guide in the search for a mitigation strategy, which should be based on trying to eliminate those fringes.


\subsection{Examples of open interferometers}
\label{sec:open_int}

\subsubsection{Asymmetric pulse timing}
\label{sec:amzi}

\begin{figure}[h]
\begin{center}
\includegraphics[width=8cm]{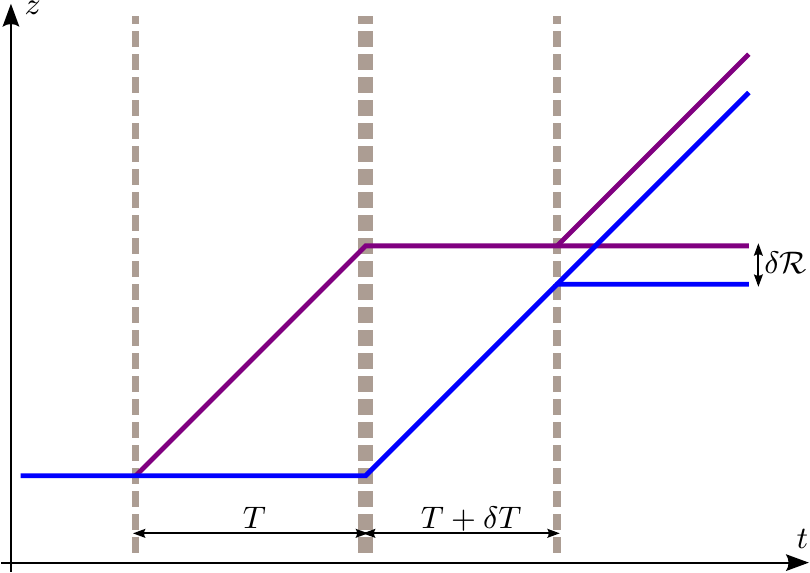}
\end{center}
\caption{Asymetric Mach-Zehnder interferometer, where a $\delta T$ difference in the time separation between the 
mirror and the two beam-splitter pulses leads to a nonvanishing relative displacement, $\delta \boldsymbol{\mathcal{R}} \neq 0$, between the trajectories of the interfering wave packets at each exit port.}
\label{fig:AMZI}
\end{figure}

We start by considering the simple example of an asymmetric Mach-Zehnder interferometer (MZI), where the time separations between the mirror pulse and the beam-splitter pulses (initial and final) differ by $\delta T$, as shown in Fig.~\ref{fig:AMZI}. In the absence of a quadratic potential there is only a position displacement between the two wave packets at the exit port and it is simply given by
\begin{equation}
\delta \boldsymbol{\mathcal{R}} = \comment{ - \mathbf{v}_\text{rec} \, \delta T } .
\end{equation}
This gives rise to a fringe pattern in the density profile with fringe spacing
\begin{equation}
\lambda_\text{fr} = \frac{2\pi\hbar\, \Delta t}{m |\delta \boldsymbol{\mathcal{R}}|} ,
\end{equation}
as observed for instance in the experiments reported in Refs.~\cite{miller05,muentinga13}.
The situation is somewhat analogous to what happens for rotations, but the displacement in that case is along the transverse directions (i.e.\ perpendicular to $\mathbf{v}_\text{rec}$); see Sec.~\ref{sec:nonaligned} below for a more detailed discussion.

\subsubsection{Gravity gradient}
\label{sec:grav_grad}

Let us consider now the effect of the gravity gradient when the dynamics is governed by a Hamiltonian like that in Eq.~\eqref{eq:hamiltonian}. Before doing so we should, however, make a couple of important remarks. Firstly, although some time will elapse in general between the last beam splitter and detection (this time should be long enough when no state labeling is employed and good spatial separation between the two ports is, thus, required), the calculation of the contrast using Eq.~\eqref{7} can be equivalently carried out right after the last beam splitter. This can be seen by rewriting the expression for the contrast as follows:
\begin{align}
C &= \Big| \big\langle \psi_\text{c} (t_\text{bs}) \big| \, \hat{U}^\dagger_0 (t,t_\text{bs}) \,
\hat{\mathcal{D}}(\delta\boldsymbol{\chi}) \,
\hat{U}_0 (t,t_\text{bs}) \, \big| \psi_\text{c} (t_\text{bs}) \big\rangle \Big|
\nonumber \\
& = \Big| \big\langle \psi_\text{c} (t_\text{bs}) \big|
\hat{\mathcal{D}}\big(\delta\boldsymbol{\chi}(t_\text{bs}) \big)
\big| \psi_\text{c} (t_\text{bs}) \big\rangle \Big|
\label{eq:contrast_bs} ,
\end{align}
where we took into account that $\big| \psi_\text{c} (t) \big\rangle = \hat{U}_0 (t,t_\text{bs})\, \big| \psi_\text{c} (t_\text{bs}) \big\rangle$ and made use in the second equality of Eq.~\eqref{B5} applied to this case together with Eq.~\eqref{eq:displacement_bs}.
Therefore, \comment{unless stated otherwise,} from now on we will always calculate the displacement $\delta\boldsymbol{\chi}$ at time $t_\text{bs}$, right after the last beam splitter.
Secondly, as mentioned in the introduction, we will neglect the effect of the gravity gradient on the evolution of the centered wave packet and use the free-particle unitary evolution operator for $\big| \psi_\text{c} (t) \big\rangle = \hat{U}_0 (t,t_0)\, \big| \psi_\text{c} (t_0) \big\rangle$ when evaluating Eq.~\eqref{7}. Using instead the full unitary evolution operator would simply add small corrections to the result for the contrast suppressed by additional powers of $\Gamma \, T^2$.

Particularizing the results of Eqs.~\eqref{eq:displacement_mzi_r}-\eqref{eq:displacement_mzi_p} to the case of a MZI with half-interferometer time $T$ and $\delta T = 0$, one obtains the following position and momentum displacements at time $t_\text{bs}$ and to first order in $\Gamma \, T^2$:
\begin{align}
\delta \boldsymbol{\mathcal{R}} &= \left(\Gamma\, T^2 \right) \mathbf{v}_\text{rec} \, T
\label{14a} , \\
\delta \boldsymbol{\mathcal{P}} &= (\Gamma\, T^2) \, m\, \mathbf{v}_\text{rec}
\label{14b} .
\end{align}
These results are valid for an arbitrary orientation, but in the next subsection and Secs.~\ref{sec:gaussian}-\ref{sec:BEC} we will concentrate on the case of aligned gravity gradients, i.e.\ the case in which the direction of the laser pulses and, hence, $\mathbf{v}_\text{rec}$ coincide with the direction of one of the principal axes of the gravity gradient tensor $\Gamma$. This in turn implies $\delta \boldsymbol{\mathcal{R}} \parallel \mathbf{v}_\text{rec}$ and $\delta \boldsymbol{\mathcal{P}} \parallel \mathbf{v}_\text{rec}$, which holds to first order but also for the exact results derived in Appendix~\ref{sec:trajectories}.
On the other hand, the case of nonaligned gravity gradients will be addressed in Sec.~\ref{sec:nonaligned}.
\begin{figure}[h]
\begin{center}
\includegraphics[width=8cm]{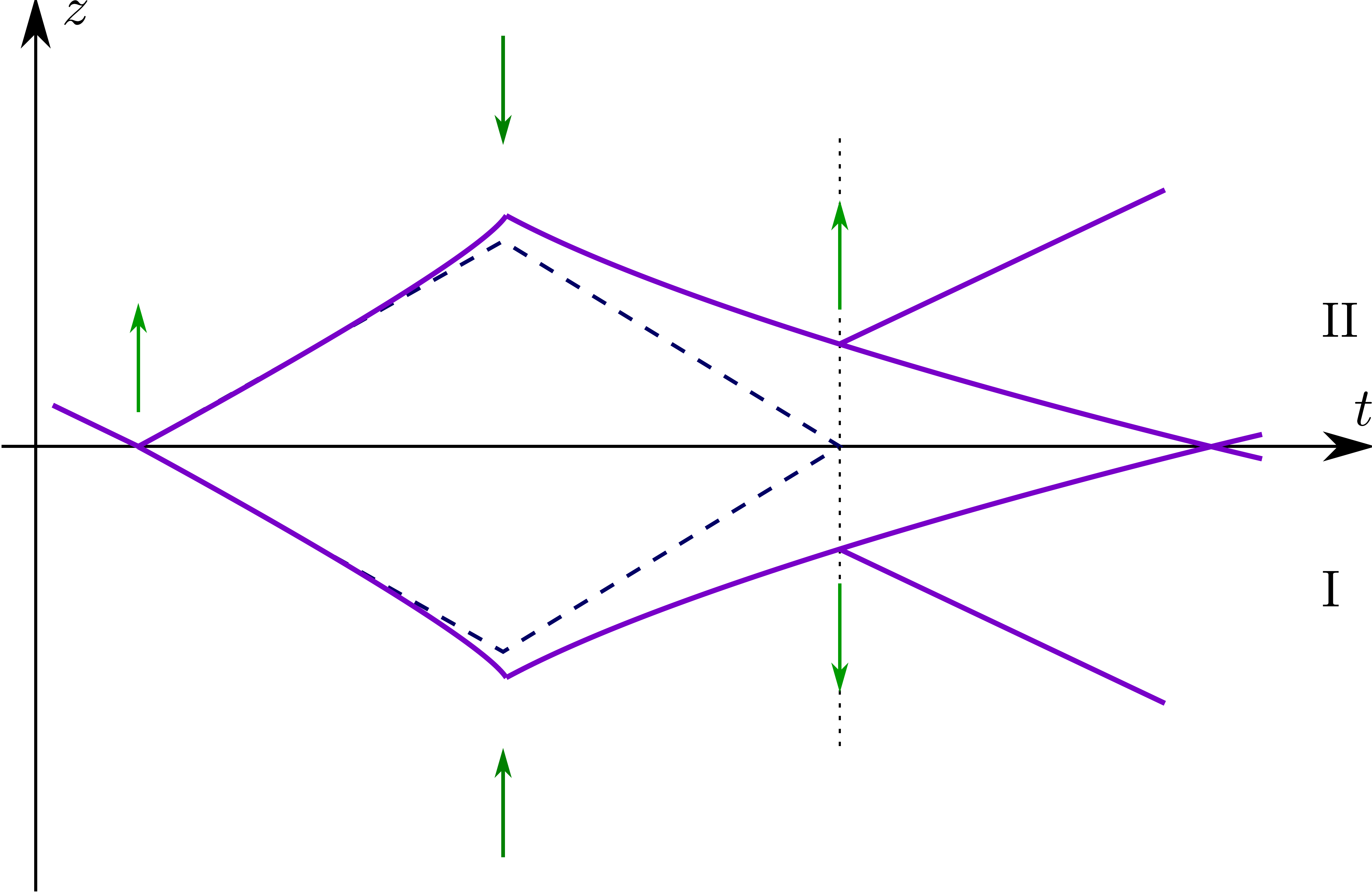}
\end{center}
\caption{Plot illustrating the effect of a quadratic potential on the wave-packet trajectories in a Mach-Zehnder interferometer (purple lines) including the momentum kicks from the laser pulses (green arrows) --- the trajectories in the absence of quadratic potential terms (dashed blue lines) are also shown for comparison. As a consequence, the two trajectories at each exit port differ in position and momentum (they have different slopes in the plot).}
\label{fig:grav_gradient}
\end{figure}

\Highlight{Note that the change of velocity associated with Eq.~\eqref{14b} is typically rather small (of the order of $10^{-4}\, \mathbf{v}_\text{rec}$ or less for half-interferometer times up to $10\,$s) and leads to negligible changes of the resonance condition given the typical finite time duration of the laser pulses (i.e.\ corresponding to energy changes much smaller than those allowed by the time-energy uncertainty relation).
On the other hand, it is interesting to point out that a vanishing final displacement in phase space, $\delta\boldsymbol{\chi}=0$, could be achieved by suitably adjusting the momentum transfer of the mirror pulse (this would lead for instance to trajectories symmetric under time inversion around the time of the mirror pulse in the case depicted in Fig.~\ref{fig:grav_gradient}). Nevertheless, for long interferometer times this would require adjusting the laser frequencies by hundreds of GHz, which is not so easy to implement. An alternative strategy which is much more flexible and easier to implement will be presented in the next subsection and studied throughout the rest of the paper.}

\subsection{Mitigation strategy}
\label{sec:strategy}

In order to mitigate the loss of contrast due to gravity gradients, one can try to combine the two examples described in the two previous subsections. Changing the timing of the last beam-splitter by a small amount $\delta T$ in the presence of a gravity gradient leads for the MZI of the previous subsection to the following position displacement to first order in $\Gamma\, T^2$ and $\delta T$:
\begin{equation}
\delta \boldsymbol{\mathcal{R}} = - \mathbf{v}_\text{rec} \, \delta T
+ \left(\Gamma\, T^2 \right) \mathbf{v}_\text{rec} \, T ,
\end{equation}
but leaves $\delta \boldsymbol{\mathcal{P}}$ invariant at that order.
It is, therefore, not possible to eliminate completely the phase-space displacement $\delta\boldsymbol{\chi}$. Nevertheless, the key insight is that by choosing appropriately $\delta T$, one can get a nonvanishing value for $\delta \boldsymbol{\mathcal{R}}$ that compensates the effect due to the nonvanishing $\delta \boldsymbol{\mathcal{P}}$ and causes the fringe spacing to be much larger than the size of the envelope. Indeed, we see from Eq.~\eqref{9} that $1/\lambda_\text{fr} \to 0$ when
\begin{equation}
\delta \boldsymbol{\mathcal{R}} - (t-t_0)\, \delta \boldsymbol{\mathcal{P}}/m = 0 
\label{eq:mitigation1},
\end{equation}
and this can be achieved for aligned gravity gradients by taking
\begin{equation}
\delta T = -(T+T_0) \big( \Gamma_{\parallel}\, T^2 \big)
\label{eq:timing} ,
\end{equation}
where $\Gamma_{\parallel}$ is the eigenvalue of the $\Gamma$ tensor along the direction of $\mathbf{v}_\text{rec}$ and we have considered a total time $t-t_0 = 2T+T_0$, with $T$ being the half-interferometer time and $T_0$ the time from $t_0$ till the first beam splitter.

The implementation of this mitigation strategy will be analyzed in more detail in the next two sections for freely evolving Gaussian states and for expanding BECs.

\section{Gaussian wave packets for free particles}
\label{sec:gaussian}

In this section we will provide a more detailed analysis specialized to Gaussian states of the questions discussed in Sec.~\ref{sec:general}. This is especially interesting because one can obtain exact results which are fairly simple and can provide additional insight on certain aspects of the loss of contrast in open interferometers and the proposed mitigation strategy. In particular, they illustrate the intuitive interpretation of our findings within a phase-space description of quantum mechanics based on the Wigner function \cite{hillary84,schleich}. Furthermore, the study of Gaussian states is also of practical interest because they can be directly applied to the description of long-time interferometry experiments using cold thermal atoms \cite{dickerson13}.
\highlight{(Provided that the system is away from quantum degeneracy and sufficiently dilute, it can be described to a good approximation in terms of one-particle distribution functions and neglecting quantum many-body correlations.)}
The case of BECs, on the other hand, will be analyzed in the next section.

\subsection{Phase-space description}
\label{sec:phase-space}

Let us consider a general pure Gaussian state as the initial state of the centered wave packet, $| \psi_\text{c} (t_0) \rangle$. Its associated Wigner function, defined in general as
\begin{equation}
W(\mathbf{x},\mathbf{p}) = \int \frac{d^3\Delta}{(2\pi\hbar)^3} \,
e^{i \mathbf{p}^\mathrm{T} \boldsymbol{\Delta}/ \hbar} \,
\big\langle \mathbf{x} - \boldsymbol{\Delta}/2 \big| \psi_\text{c} \big\rangle
\big\langle \psi_\text{c} \big| \mathbf{x} + \boldsymbol{\Delta}/2 \big\rangle
\label{eq:wigner_def},
\end{equation}
takes the form
\begin{equation}
W(\mathbf{x},\mathbf{p};t_0) = (2\pi)^{-3} (\det{\Sigma})^{-\frac{1}{2}} \,
e^{-\frac{1}{2} \boldsymbol{\xi}^\text{T} \Sigma^{-1} \boldsymbol{\xi}}
\label{17},
\end{equation}
where $\boldsymbol{\xi} = (\mathbf{x},\mathbf{p})^\text{T}$ and $\Sigma$ is the phase-space covariance matrix, which is directly related to the Weyl-ordered two-point functions:
\begin{equation}
\Sigma_{ij} = \frac{1}{2} \left\langle \hat{\xi}_i \hat{\xi}_j + \hat{\xi}_j \hat{\xi}_i \right\rangle
= \left( \begin{array}{cc} \Sigma_{xx} & \Sigma_{xp} \\
\Sigma_{xp}^\text{T} & \Sigma_{pp}
\end{array} \right)_{ij} .
\end{equation}
$\Sigma_{xx}$, $\Sigma_{xp}$ and $\Sigma_{pp}$ are $3\times3$ matrices that can be regarded as blocks of the $6\times6$ covariance matrix $\Sigma$, which is symmetric and positive definite.

In order to calculate the contrast, it is useful to note that the expectation value of a displacement operator can be obtained by computing the inverse symplectic Fourier transform of the Wigner function as follows \cite{hillary84}:
\begin{align}
\big\langle \psi_\text{c} (t) \big| \hat{\mathcal{D}}(\delta\boldsymbol{\chi})
\big| \psi_\text{c} (t) \rangle
&= \int d^3x' \int d^3p' \, W(\mathbf{x}',\mathbf{p}';t) \,
e^{-\frac{i}{\hbar} \delta \boldsymbol{\chi}^\text{T} J \, \boldsymbol{\xi}'}  \nonumber \\
&= \int d^3x \int d^3p \, W(\mathbf{x},\mathbf{p};t_0) \,
e^{-\frac{i}{\hbar} \delta \boldsymbol{\chi}_0^\text{T} J \, \boldsymbol{\xi}}
\label{21},
\end{align}
where $\delta \boldsymbol{\chi}_0 \equiv \mathcal{T} (t_0,t) \, \delta \boldsymbol{\chi}$ with the transition matrix $\mathcal{T} (t_2,t_1)$ defined in Appendix~\ref{sec:trajectories}. In the second equality we have introduced the change of variables $\boldsymbol{\xi}' = \mathcal{T} (t,t_0) \, \boldsymbol{\xi}$, used the invariance of the symplectic form $\mathcal{T}^\mathrm{T} (t,t_0) \, J \, \mathcal{T} (t,t_0)=J$ and taken into account that for quadratic potentials the Wigner function evolves exactly in the same way as a classical phase-space distribution.

From Eqs.~\eqref{21} and \eqref{17} we get the following result for the contrast defined in Eq.~\eqref{7}:
\begin{equation}
C = e^{-\frac{1}{2 \hbar^2} \delta \boldsymbol{\chi}_0^\text{T} J^\text{T} \Sigma \,
J \, \delta \boldsymbol{\chi}_0}
\label{22}.
\end{equation}
Up to a factor $1/2\hbar^2$ the exponent can be rewritten as
\begin{equation}
- \delta \boldsymbol{\mathcal{R}}_0^\text{(s)\,T} \Sigma_{pp} \,
\delta \boldsymbol{\mathcal{R}}_0^\text{(s)}
- \delta \boldsymbol{\mathcal{P}}_0^\text{T}
\Big( \Sigma_{xx} - \Sigma_{xp}\, \Sigma_{pp}^{-1} \, \Sigma_{xp}^\text{T} \Big)
\delta \boldsymbol{\mathcal{P}}_0
\label{23},
\end{equation}
with $\delta \boldsymbol{\mathcal{R}}_0^\text{(s)} = \delta \boldsymbol{\mathcal{R}}_0
- \big( \Sigma_{pp}^{-1}\, \Sigma_{xp}^\text{T} \big)\,  \delta \boldsymbol{\mathcal{P}}_0$.
In order to write $\delta \boldsymbol{\chi}_0$ in terms of $\delta \boldsymbol{\chi}$ one needs to use the transition matrix $\mathcal{T} (t_0,t)$, which is exactly given by Eq.~\eqref{A6} for time-independent gravity gradients. However, as explained in the Introduction and in Sec.~\ref{sec:open_int}, in typical cases of practical interest it is perfectly justified to treat $\Gamma$ perturbatively and neglect its effect on the evolution of the centered wave packet $| \psi_\text{c} (t) \rangle$. We will do so here, even though the exact result could be straightforwardly obtained, because the expressions are simpler and more transparent and the corrections would anyway be very small.
For free evolution we have
\begin{align}
\delta \boldsymbol{\mathcal{R}}_0 &= \delta \boldsymbol{\mathcal{R}}
- \frac{\delta \boldsymbol{\mathcal{P}}}{m} \, (t-t_0)
\label{24}, \\
\delta \boldsymbol{\mathcal{P}}_0 &= \delta \boldsymbol{\mathcal{P}}
\label{25}.
\end{align}
Thus, from Eqs.~\eqref{23}-\eqref{25} one can see that for a given $\delta \boldsymbol{\mathcal{P}}$ the contrast is maximized when $\delta \boldsymbol{\mathcal{R}}_0^\text{(s)} = 0$, which implies
\begin{equation}
\delta \boldsymbol{\mathcal{R}} = \frac{\delta \boldsymbol{\mathcal{P}}}{m} \, (t-t_0)
+ \Sigma_{pp}^{-1} \, \Sigma_{xp}^\text{T} \, \delta \boldsymbol{\mathcal{P}}
\label{26}.
\end{equation}
Note that since the exponent in Eq.~\eqref{22} is manifestly negative for any $\delta \boldsymbol{\chi}_0$ and one can always choose $\delta \boldsymbol{\mathcal{R}}_0$ so that $\delta \boldsymbol{\mathcal{R}}_0^\text{(s)}=0$ for any given $\delta \boldsymbol{\mathcal{P}}_0$, the second term in expression~\eqref{23} must be negative.
The interpretation and importance of this term will be discussed in Sec.~\ref{sec:gaussian_examples}. In this respect, it is instructive to consider the one-dimensional case, for which it becomes
\begin{equation}
- \frac{1}{2 \hbar^2} \frac{\det \Sigma}{\Sigma_{pp}} \, \delta \mathcal{P}_0^2
\leq - \frac{1}{8} \frac{\delta \mathcal{P}_0^2}{\Sigma_{pp}}
\label{eq:dP^2} ,
\end{equation}
which follows from the inequality $(\det{\Sigma}) \geq \hbar^2 / 4$. This inequality is the necessary and sufficient condition for the Gaussian phase-space distribution in Eq.~\eqref{17} to be the Wigner function of a well-defined density matrix (with $\hat{\rho}^2 \leq \hat{\rho}$); the equality holds for pure states and the strict inequality for mixed ones.
Moreover, it coincides with the Schr\"odinger uncertainty relation \cite{robertson34,trifonov02}, a somewhat stronger version of Heisenberg's uncertainty relation.

\begin{figure*}[ht]
\begin{center}
\includegraphics[width=16cm]{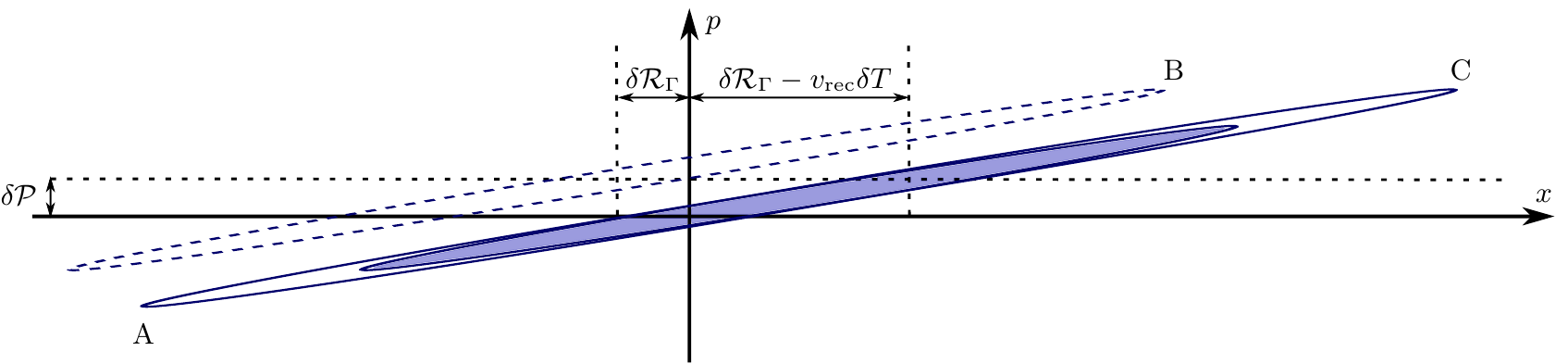}
\end{center}
\caption{Level curves for the Wigner functions (A and B) of two interfering Gaussian wave packets with a relative displacement $\delta\chi =(\delta\mathcal{R},\delta\mathcal{P})$. The loss of contrast simply corresponds to the reduced overlap of these two phase-space distributions. Provided that $\delta\mathcal{P}^2 \ll \Sigma_{pp}$, one can adjust $\delta\mathcal{R}$ with a slight timing asymmetry $\delta T$ of the laser pulses so that the two Wigner functions (A and C) are properly aligned and almost complete overlap is recovered. For simplicity a one-dimensional example is depicted.}
\label{fig:gaussian_overlap}
\end{figure*}

There is an alternative expression for Eq.~\eqref{21} as the overlap of two Wigner functions with a relative phase-space displacement $\delta \boldsymbol{\chi}$ (valid only for pure states):
\begin{align}
C &= \Big| \big\langle \psi_\text{c} (t) \big|
\hat{\mathcal{D}}(\delta \boldsymbol{\chi}) \big| \psi_\text{c} (t) \big\rangle \Big|
= \Big(\mathrm{Tr} \big[ \hat{\rho}\, \hat{\rho}_{\delta \boldsymbol{\chi}}] \Big)^\frac{1}{2}
\nonumber \\
&= (2\pi\hbar)^{\frac{3}{2}} \bigg| \int d^3x\, d^3p \, W(\mathbf{x},\mathbf{p};t)\,
W \big(\mathbf{x}-\delta\boldsymbol{\mathcal{R}},\mathbf{p}-\delta\boldsymbol{\mathcal{P}};t \big) \bigg|^{\frac{1}{2}}
\label{27},
\end{align}
where the density matrices $\hat{\rho}$ and $\hat{\rho}_{\delta \boldsymbol{\chi}}$ correspond respectively to the states $\big| \psi_\text{c} (t) \big\rangle$ and $\hat{\mathcal{D}}(\delta \boldsymbol{\chi}) \big| \psi_\text{c} (t) \big\rangle$.
Making use of Eq.~\eqref{27} to interpret the contrast as the overlap between the two shifted Wigner functions can be quite illuminating and provides an intuitive explanation, within the phase-space formulation of quantum mechanics, for the loss of contrast and the reason why the mitigation strategy works. Indeed, as shown in Fig.~\ref{fig:gaussian_overlap}, a nonvanishing relative displacement leads in general to decreasing overlap between the displaced Wigner functions (for simplicity we focus again on the one-dimensional case). Furthermore, it is clear that for fixed $\delta\mathcal{P}$ the overlap is maximized when $\delta\mathcal{R}$ is chosen so that the two displaced Wigner functions are aligned%
\footnote{If we were considering displacement variations perpendicular to the principal axis of the level-curve ellipses, maximum overlap would be achieved for perfect alignment. However, since we consider displacements with fixed $\delta \mathcal{P}$, there is in general some change of the displacement component along the principal axis too. \comment{This implies a deviation in the alignment condition for contrast maximization which becomes smaller for larger squeezing \cite{zeller14a}.}}.
Taking into account that $\Sigma_{pp}^{-1} \, \Sigma_{xp}^\text{T}$ characterizes the ``tilt'' in phase space of the initial Wigner function and considering its free evolution, one can easily obtain condition~\eqref{26} for maximum contrast. It is also clear that substantial overlap will be achieved as long as $\delta \mathcal{P} \ll (\Sigma_{pp})^{1/2}$.
Finally, it should be noted that for generic (non-Gaussian) pure states evolution for a sufficiently long time will lead to significantly squeezed Wigner functions and arguments analogous to those just presented for Gaussian states will still hold at the qualitative level.

The result in Eq.~\eqref{26} agrees at late times with the late-time result~\eqref{eq:mitigation1} derived in Sec.~\ref{sec:general}. Furthermore, the second term on the right-hand side of Eq.~\eqref{26}, which is proportional to $\Sigma_{xp}$ and suppressed by a factor $1/\Delta t$ compared to the first one, would have also been obtained in Sec.~\ref{sec:general} if we had considered spatially dependent phase contributions from $\tilde{\psi} (m\, \mathbf{x}/\Delta t,t_0)$, which are always present for $\Sigma_{xp} \neq 0$.

\subsection{Pure \emph{versus} mixed states}
\label{sec:gaussian_mixed}

There is an interesting generalization of the previous results to the case of mixed states, as we explain below. First, one writes the initial density matrix as
\begin{equation}
\hat{\rho}(t_0) = \hat{\mathcal{D}}(\boldsymbol{\chi}_0)\, \hat{\rho}_\text{c} (t_0) \,
\hat{\mathcal{D}}^\dagger (\boldsymbol{\chi}_0)
\label{eq:rho},
\end{equation}
so that $\mathrm{Tr} \big[ \hat{\boldsymbol{\xi}}\, \hat{\rho}_\text{c} (t_0) \big] = 0$. Next, we express $\hat{\rho}_\text{c} (t_0)$ as an incoherent mixture of pure states and write the density matrix for the initial state in the following way:
\begin{equation}
\hat{\rho}_\text{c}(t_0) = \sum_j p_j \big| \psi^{(j)}(t_0) \big\rangle \big\langle \psi^{(j)}(t_0) \big|
\label{eq:rho_c}
\end{equation}
with $0 \leq p_j \leq 1$ and $\sum_j p_j = 1$.
One can then apply the results obtained for pure states to each state $\big| \psi^{(j)}(t_0) \big\rangle$ and finally take the average over the whole ensemble. In doing so, it is convenient not to impose the condition that the expectation value of the operator $\boldsymbol{\hat{\xi}}$ vanishes for centered states (this was done to specify them completely, 
but we needn't have done so and all the results derived for pure states would still hold since no further use of the condition was made in any of the calculations and derivations). One can then use this freedom to choose $\big| \psi^{(j)}(t_0) \big\rangle$ as centered states.
With this choice the term $C \cos \delta\phi$ appearing in Eq.~\eqref{6} corresponds for each pure state of the ensemble to the real part of
\begin{equation}
C^{(j)} e^{i \delta \phi^{(j)}} \equiv
\big\langle \psi^{(j)} (t) \big| \hat{\mathcal{D}}(\delta\boldsymbol{\chi})
\big| \psi^{(j)} (t) \big\rangle \, e^{i \delta \Phi}
\label{eq:contrast_j}
\end{equation}
where $C^{(j)}$ is defined as the modulus of the expression on the right-hand side (recall that $\delta \phi^{(j)}$ contains a possible additional phase from the expectation value of the displacement operator).
The relative displacement $\delta\boldsymbol{\chi}$ is the same for all members of the ensemble, and so is $\delta\Phi$, which depends on the initial displacement $\boldsymbol{\chi}_0$ appearing in Eq.~\eqref{eq:rho}, but that is also the same for every member.
After taking the average over the whole ensemble, one is naturally led to the following quantity: 
\begin{equation}
C\, e^{i \delta\phi} = \sum_j p_j \, C^{(j)} e^{i \delta \phi^{(j)}} 
\end{equation}
whose real part characterizes the oscillations of the total particle number at the exit port for the whole ensemble, as given by Eq.~\eqref{6}. In particular, for the contrast one gets
\begin{equation}
C = \Big| \sum_j p_j \, C^{(j)} e^{i \delta \phi^{(j)}} \Big|
\leq \sum_j p_j \, C^{(j)}
\label{eq:contrast_ensemble}
\end{equation}
where one has a strict inequality unless all the phases $\delta \phi^{(j)}$ are equal. Thus, we see that \emph{dephasing} between the different members of the ensemble can lead to a further reduction of the contrast for the oscillations of the total particle number at each port as a function of $\delta\phi$, even when there is hardly any loss of contrast for each member of the ensemble.

It is clear from Eq.~\eqref{eq:contrast_ensemble} that the result for pure states in Eq.~\eqref{7} admits the following natural generalization to mixed states:
\begin{equation}
C = \Big| \mathrm{Tr} \big[ \hat{\mathcal{D}}(\delta\boldsymbol{\chi})\, \hat{\rho}_\text{c} (t) \big]\Big| ,
\end{equation}
which in turn implies that Eq.~\eqref{21}, in terms of the Wigner function, can be directly applied to mixed states as well. [The Wigner function for mixed states is defined by replacing $\big| \psi_\text{c} \big\rangle \big\langle \psi_\text{c} \big|$ with $\hat{\rho}_\text{c}$ in Eq.~\eqref{eq:wigner_def}.] In contrast, Eq.~\eqref{27}, which depends nonlinearly on the Wigner function, only holds for pure states.
It should also be noted that if we had kept the condition of vanishing expectation value of the operator $\boldsymbol{\hat{\xi}}$ for the centered states, this would have led to a different initial displacement $\boldsymbol{\chi}_0^{(j)}$ for each pure state $\big| \psi^{(j)}(t_0) \big\rangle$ and, hence, a different $\delta \Phi^{(j)}$ too. However, this would be exactly compensated by the expectation value of the displacement operator with respect to the new centered states $\big| \psi_\text{c}^{(j)}(t) \big\rangle$ so that $\delta \phi^{(j)}$ would remain unchanged.

Finally, it should be emphasized that so far the whole discussion and conclusions about mixed states, including the validity of Eq.~\eqref{21}, were completely general and not just restricted to Gaussian states. When particularizing to Gaussian density matrices, one can further conclude that Eq.~\eqref{22}, derived for pure Gaussian states in the previous subsection, can be directly applied to mixed states as well.

\subsection{Quantitative examples}
\label{sec:gaussian_examples}

\begin{figure}[h]
\begin{center}
\includegraphics[width=8.5cm]{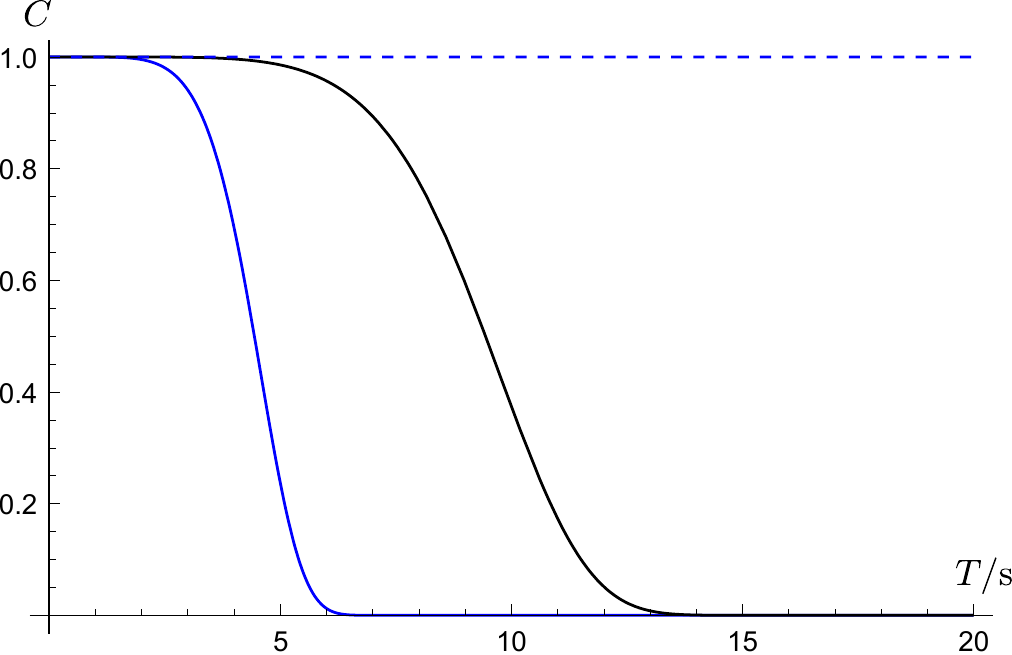}
\end{center}
\caption{Integrated contrast as a function of the half-interferometer time for pure Gaussian states with $\Sigma_{zz}=(100\, \mu\text{m})^2$, $\Sigma_{pp}$ corresponding to $T_\text{eff} = 1\, \text{nK}$ (blue curves) and the corresponding nonvanishing value of $\Sigma_{zp}$. One can see a substantial loss of contrast caused by an aligned gravity gradient with $\Gamma_{zz} = 3 \cdot 10^{-6}\, \text{s}^{-2}$ (continuous blue line) and how this improves dramatically when the mitigation strategy is employed (dashed line). The result for a 10\% residual gravity gradient is also shown for comparison (black curve).}
\label{fig:gaussian_pure}
\end{figure}

In order to illustrate how effective our proposed mitigation strategy can be, we consider the effects of an aligned gravity gradient with $\Gamma_{zz} = 3 \cdot 10^{-6}\, \text{s}^{-2}$, comparable to that on the surface of Earth, and a \emph{pure} Gaussian state for $^{87}\text{Rb}$ atoms with $\Sigma_{zz} = (100\, \mu m)^2$ and a momentum width along this direction corresponding to an effective temperature $T_\text{eff} = 1\, \text{nK}$ (implying a velocity width of $0.3\, \text{mm/s}$). This temperature, which has already been successfully implemented in interferometry experiments employing the so-called delta-kick cooling technique \cite{muentinga13}, is much less demanding than the requirement of $T_\text{eff} \approx 70\, \text{pK}$ in current plans for STE-QUEST, mainly driven by the loss of contrast due to gravity gradients when no mitigation strategy is employed.
(For simplicity we assume vanishing nondiagonal elements involving the $z$ direction and the perpendicular directions within the matrix $\Sigma$, so that those transverse directions can be trivially integrated out and the example reduces to a one-dimensional problem. Moreover, we assume that the initial conditions are specified right before the first beam splitter, i.e.\ we take $T_0 = 0$.)
In Fig.~\ref{fig:gaussian_pure} one can see how the gravity gradient leads to a substantial loss of contrast for $T=5\, \text{s}$ of half-interferometer time, whereas the mitigation strategy enables an extension beyond $T=20\, \text{s}$ with virtually no loss of contrast. This requires perfect alignment and perfect knowledge of the gravity gradient, but even the rather conservative assumption that this can only be achieved at the 10\% level, so that a residual 10\% of the original gravity gradient remains, extends the possible half-interferometer time to close to $T=10\, \text{s}$.
\Highlight{Note that determining the gravity gradient with a 10\% accuracy is quite feasible since the effects of the local topography and nearby objects are typically of that order at most. Moreover, commercially available gradiometers can provide measurements of the gravity gradient with 0.1\% accuracy, and the contribution of local masses can also be modeled with sufficient accuracy, for instance the payload in a space mission.}

\begin{figure}[h]
\begin{center}
\includegraphics[width=8.5cm]{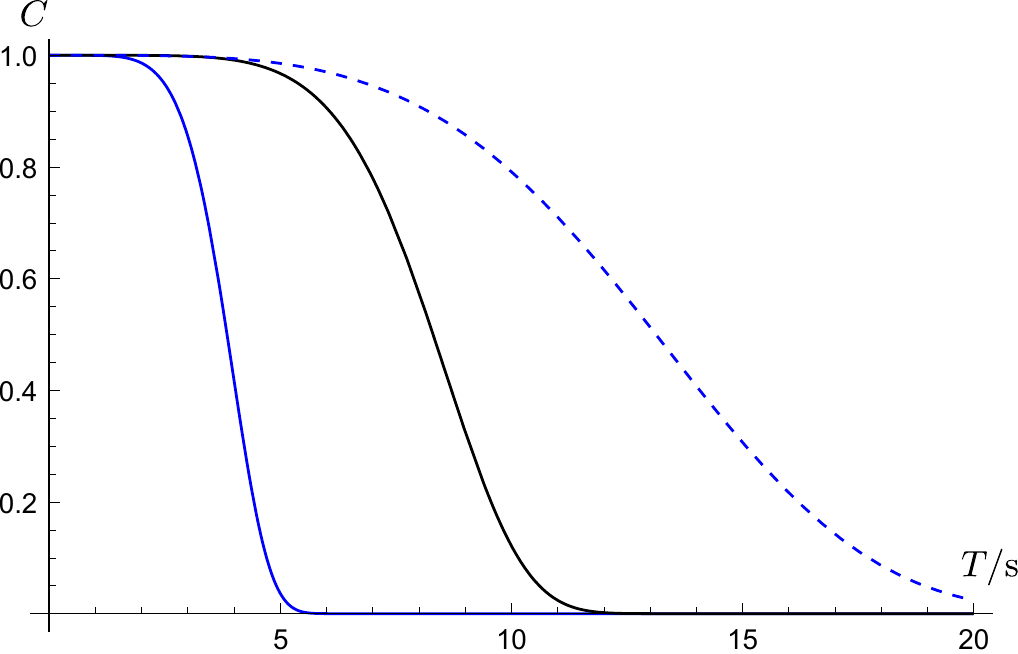}
\end{center}
\caption{Same as Fig.~\ref{fig:gaussian_pure} but for a mixed Gaussian state with $\Sigma_{zp}=0$, $\Sigma_{zz}=(141\, \mu\text{m})^2$ and $\Sigma_{pp}$ corresponding to $T_\text{eff} = 2\, \text{nK}$, which can be understood as an incoherent ensemble of pure states like the one considered there. The faster decrease of the integrated contrast compared to Fig.~\ref{fig:gaussian_pure} (especially when the mitigation strategy is employed) is due to the dephasing between the different members of the ensemble.}
\label{fig:gaussian_mixed}
\end{figure}

In the previous example $\Sigma_{zp}$ had a nonvanishing value which was determined (up to a sign) by the requirement that $(\det{\Sigma}) = \hbar^2 / 4$ for pure Gaussian states. As a second example we will consider a Gaussian \emph{mixed} state under the same conditions but with $\Sigma_{zp} = 0$ as well as $\Sigma_{zz}=(141\, \mu\text{m})^2$ and $\Sigma_{pp}$ corresponding to $T_\text{eff} = 2\, \text{nK}$.
(For this choice the mixed state can be regarded as an incoherent ensemble of pure states like that in the first example.)
As can be seen in Fig.~\ref{fig:gaussian_mixed}, the mitigation strategy is somewhat less effective in this case and the contrast is reduced almost completely for $T \gtrsim 10\, \text{s}$. This has a simple intuitive explanation in terms of the dephasing between the different members of the ensemble of pure states discussed in Sec.~\ref{sec:gaussian_mixed} and leading to the inequality in Eq.~\eqref{eq:contrast_ensemble}. In fact, the Wigner function associated with a mixed Gaussian state can be written as a convolution of identical Wigner functions for pure Gaussian states but centered at different points in phase space, and whose centers follow a Gaussian distribution. Such a decomposition is not unique (there are infinitely many of them), but for the second example the Wigner functions in this convolution \comment{can be chosen to coincide (except for their centers) with the pure state in our first example \cite{roura14b}.} The loss of contrast and the effect of the mitigation strategy for each member of the ensemble are then the same as in that example. However, there is an additional reduction of contrast due to relative dephasing between the members of the ensemble and this remains even when the mitigation strategy is employed.

We finish this section by discussing the role and interpretation of the term quadratic in $\delta \boldsymbol{\mathcal{P}}$ appearing in Eq.~\eqref{23}, which is entirely responsible for any remaining contrast reduction when the mitigation strategy is used. 
For pure states this is due to the lack of full overlap between the two displaced Wigner functions, even after they have been aligned, because  $\delta \boldsymbol{\mathcal{P}}$ is fixed. This can be easily illustrated for the one-dimensional case, where the ratio between the relative displacement of the aligned Wigner functions and their width along the alignment direction equals $\delta \mathcal{P} / \big( 2\, \sqrt{\Sigma_{pp}} \big)$, and minus one half the square of this quantity [as in Eq.~\eqref{eq:dP^2}] gives, when exponentiated, the overlap between two such Gaussian Wigner functions. In position representation this is directly related to the partial overlap of the envelopes of the two wave packets,
whereas the phase-space alignment essentially corresponds to the lack of oscillations in the probability density of the wave-packet superposition.
Substituting Eq.~\eqref{14b} into Eq.~\eqref{eq:dP^2}, one gets a relation for pure states from which one can immediately see how small this term is in typical situations of practical interest (the smaller, the better since it appears as a negative exponent):
\begin{equation}
\frac{1}{8} \frac{\delta \mathcal{P}^2}{\Sigma_{pp}}
= \frac{1}{8} \bigg[  \big( \Gamma_{zz} \, T^2 \big)
\frac{v_\text{rec}}{\sigma_v} \bigg]^2
\label{eq:v_width}.
\end{equation}
Thus, for $T=10\, \text{s}$ and typical values of the gravity gradient on Earth's surface this term is very small all the way down to $\sigma_v \sim 10^{-3}\, v_\text{rec}$, which corresponds to an extremely narrow velocity distribution
(and $T_\text{eff} \sim 1\, \text{pK}$ for $^{87} \text{Rb}$).
On the other hand, as seen in Eq.~\eqref{eq:dP^2}, for mixed states the result in Eq.~\eqref{eq:v_width} is enhanced by a factor $4\, (\det \Sigma) /\hbar^2 > 1$ compared to a pure state with the same $\Sigma_{pp}$, and this characterizes the strength of the relative dephasing between the members of the ensemble mentioned above.

\section{Expanding  BEC\lowercase{s}} 
\label{sec:BEC}

\subsection{Analytical results within the scaling approach}
\label{sec:BEC_analytical}

Throughout this section we will consider an expanding BEC evolving through an interferometer sequence after being released from a trap \cite{torii00,muentinga13}, and will assume that it is dilute enough by the time the first beam-splitter pulse is applied so that the effect of interatomic interactions can be ignored when describing the evolution throughout the interferometer. 
This means that during that period the nonlinear terms in the Gross-Pitaevskii equation \eqref{eq:GP} governing the dynamics of the condensate can be neglected and it reduces to the Schr\"odinger equation for single atoms.
One can then directly apply the methods and results of Sec.~\ref{sec:general} by taking an initial state $|\psi_\text{c} (t_1) \rangle$ which is obtained after evolving the wave function of the condensate with the Gross-Pitaevskii equation up to a time when nonlinearities can be neglected.
In this context, the generalization of the Thomas-Fermi approximation to time-dependent potentials which naturally arises within the so-called scaling approach, briefly reviewed in Appendix~\ref{sec:BEC_evolution}, provides a simple analytical result for the wave function of an expanding BEC after release from the trap:
\begin{equation}
\psi_\text{c} (\mathbf{x},t) \approx \big( N \det \Lambda(t) \big)^{-\frac{1}{2}} \, e^{i \beta(t)}
\, e^{i \frac{m}{2\hbar} \mathbf{x}^\text{T} \dot{\Lambda}\Lambda^{-1} \mathbf{x}}\, 
\psi_\text{TF} (\Lambda^{-1}\mathbf{x})
\label{eq:scaling_approx} ,
\end{equation}
where the different objects appearing in this expression are defined in Eqs.~\eqref{eq:beta}, \eqref{eq:lambda}-\eqref{eq:initial_conds} and \eqref{eq:psi_lambda}.
Furthermore, it also describes correctly the free evolution at arbitrarily late times, as shown in Appendix~\ref{sec:scaling_vs_free}.
Therefore, one can directly employ Eq.~\eqref{eq:scaling_approx} without the need to match the solution for the early nonlinear regime to the free evolution governed by the Schr\"odinger equation at late times.
\comment{Note that since $|\psi_\text{TF}|^2$ is normalized to the number of atoms in the BEC, we have divided by $\sqrt{N}$ so that $\psi_\text{c}$ agrees with the single-particle normalization employed in Secs.~\ref{sec:general}-\ref{sec:gaussian}.}

Substituting the scaling solution~\eqref{eq:scaling_approx} into Eq.~\eqref{eq:prob_density1}, one obtains the following result for the probability density, analogous to Eq.~\eqref{eq:prob_density2}:
\begin{align}
\big|\psi_\text{I} (\mathbf{x},t) \big|^2 \approx \, & \frac{1}{4 N \det \Lambda} \,
\Big| 1 + e^{i \delta\phi} e^{\frac{i}{\hbar} \left( \delta \boldsymbol{\mathcal{P}}
- m \dot{\Lambda}\Lambda^{-1} \delta \boldsymbol{\mathcal{R}} \right)^\text{T}
\mathbf{x}} \Big|^2 \nonumber \\
& \times \big|\psi_\text{TF} (\Lambda^{-1}\mathbf{x}) \big|^2
\label{eq:BEC1}
\end{align}
where we have assumed that the size of the envelope (the rescaled Thomas-Fermi radius) is much larger than $\delta \boldsymbol{\mathcal{R}}$ and have neglected the relative shift between the envelopes.
The mitigation strategy is completely analogous to that already discussed in Sec.~\ref{sec:strategy} for the free-particle case. From Eq.~\eqref{eq:BEC1} we see that the limit of large fringe spacing (larger than the size of the envelope) corresponds to the condition
\begin{equation}
\delta \boldsymbol{\mathcal{P}}
- m \, \dot{\Lambda}\Lambda^{-1} \delta \boldsymbol{\mathcal{R}} = 0
\label{eq:BEC_mitigation}.
\end{equation}
This can be achieved with a small shift $\delta T$ in the timing of the last beam-splitter pulse, which can be easily determined by substituting Eqs.~\eqref{14a}-\eqref{14b} into Eq.~\eqref{eq:BEC_mitigation} and gives a result similar to Eq.~\eqref{eq:timing}.

At late times $\Lambda_{ii} \approx c_i + b_i \, (t-t_0)$ for $i=1,2,3$ \comment{in the basis where the original trap diagonalizes,} and we get the following expansion:
\begin{equation}
\big( \dot{\Lambda} \Lambda^{-1} \big)_{ii} = \frac{1}{\Delta t} \bigg[ 1 - \frac{c_i}{b_i} \frac{1}{\Delta t}
+ O \big( 1/\Delta t^2 \big) \bigg]
\label{eq:late-time_A} .
\end{equation}
Thus, to lowest order $\dot{\Lambda}\Lambda^{-1} \approx (1/\Delta t) \, \mathbb{1}$ and the late-time limit of Eq.~\eqref{eq:BEC1} agrees with the result in Eq.~\eqref{eq:prob_density2}.
Specifically, the terms of order $1$ and $1/\Delta t$ in Eq.~\eqref{eq:BEC_mitigation} and the exponent of Eq.~\eqref{eq:BEC1} coincide with those in Eq.~\eqref{9} of Sec.~\ref{sec:contrast_loss}.
Moreover, if one takes an initial time $t_0$ when the expansion is already in the linear regime, the term of order $1/\Delta t^2$ on the right-hand side of Eq.~\eqref{eq:late-time_A} gives rise to a contribution in Eq.~\eqref{eq:BEC_mitigation} analogous to the term proportional to $\Sigma_{pp}^{-1} \, \Sigma_{xp}^\text{T}$ in Eq.~\eqref{26} for the Gaussian case. This term is a consequence of having $\dot{\Lambda}(t_0) \neq 0$ and would have also been obtained in Sec.~\ref{sec:contrast_loss} if the existence of a spatially dependent imaginary part of $\tilde{\psi} (m\, \mathbf{x}/\Delta t,t_0)$, which it implies, had been taken into account there.

Before providing several quantitative examples in the next subsection, we derive some useful analytic expressions for the \comment{integrated contrast} of interferometers based on expanding BECs.
It can be calculated by substituting the scaling solution~\eqref{eq:scaling_approx} into Eq.~\eqref{10}.
If one neglects $\delta \boldsymbol{\mathcal{R}}$ compared to the size of the envelope, which is a very good approximation in most cases of practical interest (as illustrated by Fig.~\ref{fig:BEC} and further discussed below), one gets the following result, analogous to Eq.~\eqref{11}:
\begin{equation}
C \approx \frac{1}{N} \left| \int d^3 \zeta
\cos\left[ \big( \delta \boldsymbol{\mathcal{P}}^\mathrm{T} \Lambda
- m\, \delta \boldsymbol{\mathcal{R}}^\mathrm{T} \dot{\Lambda} \big) \boldsymbol{\zeta} /\hbar \right]
\psi_\text{TF}^2 (\boldsymbol{\zeta}) \right|
\label{eq:contrast_BEC1} ,
\end{equation}
where $\boldsymbol{\zeta}$ are the rescaled coordinates defined in Eq.~\eqref{eq:zeta} and it has been taken into account that $\psi_\text{TF} (\boldsymbol{\zeta})$, given by Eq.~\eqref{eq:psi_lambda}, is a real and even function. We have also made use of the equality $\Lambda^\mathrm{T} \dot{\Lambda}  = \dot{\Lambda}^\mathrm{T} \Lambda$, proved at the end of Appendix~\ref{sec:scaling}.

For $\mathbf{v}_\text{rec}$ aligned with a principal axis of both the trapping potential and the gravity gradient tensor $\Gamma$, one can choose a Cartesian coordinate system where $\Lambda = \text{diag} \big( \Lambda_{\parallel}, \Lambda_{\perp}^{(1)}, \Lambda_{\perp}^{(2)} \big)$ with the longitudinal direction $\parallel$ corresponding to the direction of $\mathbf{v}_\text{rec}$.
Integrating over the transverse directions, Eq.~\eqref{eq:contrast_BEC1} becomes
\begin{equation}
C \approx \frac{15}{16} \left| \int_{-R_\text{TF}^\parallel}^{R_\text{TF}^\parallel}
\frac{d \zeta_\parallel}{R_\text{TF}^\parallel}
\cos \left( \frac{2\pi}{\bar{\lambda}_\text{fr}} \zeta_{\parallel} \right)
\left[ 1- \left(\frac{\zeta_\parallel}{R_\text{TF}^\parallel} \right)^{\!\! 2} \right]^2 \right|
\label{eq:contrast_BEC1b} ,
\end{equation}
where $R_\text{TF}^\parallel$ is the Thomas-Fermi radius along the longitudinal direction and $2\pi/\bar{\lambda}_\text{fr} = \big( \delta\mathcal{P}\, \Lambda_{\parallel} - m\, \delta \mathcal{R}\, \dot{\Lambda}_{\parallel} \big)/\hbar$ with $\bar{\lambda}_\text{fr}$ being the fringe spacing in the rescaled coordinates.
The result in Eq.~\eqref{eq:contrast_BEC1b} can be easily generalized to nonaligned gravity gradients and traps. First, one determines $\lambda_\text{fr}$ and the longitudinal direction from the modulus and direction of the vector contracted with $\boldsymbol{\zeta}$ within the argument of the cosine in Eq.~\eqref{eq:contrast_BEC1} (note that the matrix $\Lambda(t)$ is not necessarily diagonal anymore). Next, one integrates the two transverse directions, which results in the 1-D marginal distribution of the density distribution $\psi_\text{TF}^2 (\boldsymbol{\zeta})$. As shown in Ref.~\cite{arnold}, the form of this marginal distribution is the same as that appearing in Eq.~\eqref{eq:contrast_BEC1b} even for
nonaligned traps: one simply needs to replace the square of the Thomas-Fermi radius $( R_\text{TF}^\parallel )^2$ with the $\Sigma_{\parallel \parallel}$ component of the matrix $\Sigma$ characterizing the ground state in Eq.~\eqref{eq:psi_lambda}. Thus, Eq.~\eqref{eq:contrast_BEC1b} is also applicable to the general case with this simple replacement.

\subsection{Quantitative examples}
\label{sec:BEC_examples}

In this subsection we provide several quantitative examples of contrast loss due to gravity gradients and the outcome of the mitigation strategy for the case of expanding BECs. All our results are based on the approximate scaling solution~\eqref{eq:scaling_approx}.

\begin{figure}[h]
\begin{center}
\includegraphics[width=8.5cm]{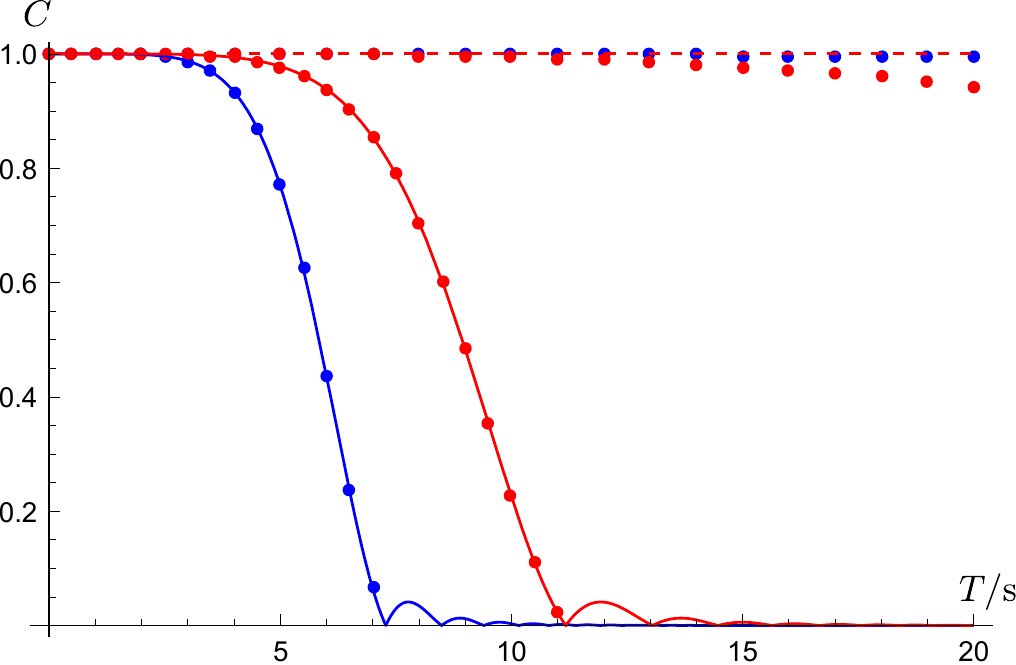}
\end{center}
\caption{Integrated contrast as a function of the half-interferometer time for an expanding BEC of $^{87}\text{Rb}$ atoms with $T_\text{eff} = 1\, \text{nK}$ (blue) and $T_\text{eff} = 70\, \text{pK}$ (red). The loss of contrast is due to a gravity gradient $\Gamma_\parallel = 3 \cdot 10^{-6}\, \text{s}^{-2}$. The curves for the cases with mitigation strategy (dashed lines) and without (continuous lines) correspond to the analytic computation neglecting the shift between the envelopes. For comparison, the dots are the result of numerically evaluating the integral along the longitudinal direction when the relative shift of the envelopes is taken into account.}
\label{fig:BEC}
\end{figure}

Similarly to what was done in Sec.~\ref{sec:gaussian_examples} for Gaussian states, we will consider a gravity gradient with $\Gamma_\parallel = 3 \cdot 10^{-6}\, \text{s}^{-2}$ and aligned with a principal axis of the trap potential and $\mathbf{v}_\text{rec}$.
In addition, we will characterize our BEC of $^{87}\text{Rb}$ atoms at some time $t_0$ when most of the repulsive interaction energy has been converted into kinetic energy and the expansion dynamics has entered the linear regime, with $\Lambda(t) = C + B\, (t-t_0)$. Furthermore, since the result for the contrast in Eq.~\eqref{eq:contrast_BEC1b} only depends on the product $R_\text{TF}^\parallel \, \Lambda_\parallel (t)$ (and its time derivative), one can entirely parametrize the system in terms of $R_\text{TF}^\parallel \, \Lambda_\parallel (t_0)$ and $R_\text{TF}^\parallel \, \dot{\Lambda}_\parallel (t_0)$, which determine the size of the condensate and the width of its velocity distribution at that time. The corresponding results are plotted in Fig.~\ref{fig:BEC} (with and without the mitigation strategy) for an initial half-width $R_\text{TF}^\parallel \, \Lambda_\parallel (t_0) = \comment{60\, \mu\text{m}}$ and two different widths of the velocity distribution: $0.3\, \text{mm/s}$ and \comment{$0.08\, \text{mm/s}$} (with associated effective temperatures of $1\, \text{nK}$ and $70\, \text{pK}$).

From Eq.~\eqref{eq:contrast_BEC1} and as seen in Fig.~\ref{fig:BEC}, it is clear that the loss of contrast due to an aligned gravity gradient is entirely removed by the mitigation strategy when the shift between the envelopes is neglected. Indeed, when using the approximate scaling solution~\eqref{eq:scaling_approx}, the only source of contrast loss is the relative shift between the envelopes of the two wave packets in the superposition at each exit port. Thus, in order to capture this effect, which is small but eventually appears at sufficiently late times, one needs to consider the result of substituting the scaling solution~\eqref{eq:scaling_approx} into Eq.~\eqref{10} without neglecting the shift between the two envelopes:
\begin{align}
C \approx \frac{1}{N} \bigg| \int & d^3 \zeta \,
\cos \Big[\big( \delta \boldsymbol{\mathcal{P}}^\mathrm{T} \Lambda
- m\, \delta \boldsymbol{\mathcal{R}}^\mathrm{T} \dot{\Lambda} \big)\, \boldsymbol{\zeta}\,
/ \hbar \Big] \,
\nonumber \\
& \times
\psi_\text{TF} \big( \boldsymbol{\zeta} + \Lambda^{-1} \delta \boldsymbol{\mathcal{R}}/2 \big)  \psi_\text{TF} \big(\boldsymbol{\zeta} - \Lambda^{-1} \delta \boldsymbol{\mathcal{R}}/2
\big) \bigg|
\label{eq:contrast_BEC2} ,
\end{align}
which is an exact result for the centered wave function $\psi_\text{c} (\mathbf{x},t)$ given by Eq.~\eqref{eq:scaling_approx}. In deriving Eq.~\eqref{eq:contrast_BEC2} we have taken again into account that $\psi_\text{TF} ( \boldsymbol{\zeta})$ is a real and even function.

For aligned gravity gradients and traps one can proceed similarly to what was done above to obtain Eq.~\eqref{eq:contrast_BEC1b}. While the integrals with respect to the transverse coordinates in Eq.~\eqref{eq:contrast_BEC2} can also be performed analytically, in this case there is no simple analytical result for the integral with respect to the longitudinal coordinate $\zeta_{\parallel}$ due to the more complicated functional dependence of $\psi_\text{TF} \big( \boldsymbol{\zeta} + \Lambda^{-1} \delta \boldsymbol{\mathcal{R}}/2 \big) \, \psi_\text{TF} \big(\boldsymbol{\zeta} - \Lambda^{-1} \delta \boldsymbol{\mathcal{R}}/2 \big)$ compared to $\psi_\text{TF}^2 ( \boldsymbol{\zeta})$.
This last integral can be evaluated numerically, which was done to obtain the results represented as dots in Fig.~\ref{fig:BEC}. Given their agreement with the continuous curves, the figure shows that neglecting the relative shift of the envelopes is indeed a very good approximation for situations of practical interest, as already discussed at the end of Sec.~\ref{sec:gaussian_examples}. One can also see that (within the time-dependent Thomas-Fermi approximation) this relative shift is the only source of contrast loss due to gravity gradients when the mitigation strategy is employed. For the examples plotted in Fig.~\ref{fig:BEC} the effect is very small and can only start to be appreciated for very long half-interferometer times. Moreover, it is slightly more significant for lower $T_\text{eff}$ because the size of the envelopes at late times is smaller in that case.


\section{Rotations and nonaligned gravity gradients}
\label{sec:nonaligned}

Besides asymmetric pulse timing and gravity gradients, another effect leading to open MZIs are rotations. Given an otherwise closed MZI in a uniform (and possibly time-dependent) force field, changes in the direction of the wave vector $\mathbf{k}_i$ for the different laser pulses will lead in general to
\begin{equation}
\left(\begin{array}{c}\delta \boldsymbol{\mathcal{R}} \\ \delta \boldsymbol{\mathcal{P}}\end{array} \right)
= \sum_{i=1}^n  \mathcal{T} (t,t_i)
\left( \begin{array}{c} \mathbf{0} \\ \comment{\delta( \varepsilon_i  \, \hbar \mathbf{k}_i )} \end{array} \right)
\neq \mathbf{0} .
\end{equation}
Since we are working in the nonrotating frame as far as the dynamics of the atoms is concerned, 
the effect of rotations is to rotate the effective momentum transfer associated with each laser pulse \cite{kleinert14a}. (It can also cause a rotation of $\mathbf{g}(t)$, but that does not alter the displacement, which is independent of $\mathbf{g}(t)$ as explained in Appendix~\ref{sec:trajectories}.)
For example, for a MZI a uniform angular velocity $\boldsymbol{\Omega}$ produces the following displacements after the last beam splitter:
\begin{align}
\delta \boldsymbol{\mathcal{R}}
&= 2\, (\mathbf{v}_\text{rec}\, T) \times(\boldsymbol{\Omega}\, T)
+ O \big( (\Omega\, T)^2 \big) , \\
\delta \boldsymbol{\mathcal{P}} &
= O \big( (\Omega\, T)^2 \big) .
\end{align}
Thus, to lowest order in $\Omega\, T$ we have a position displacement perpendicular to $\mathbf{v}_\text{rec}$ and no momentum displacement. Once again this leads to the appearance of a fringe pattern in the density profile at each exit port and a loss of contrast in the oscillations of the integrated particle number as described in Sec.~\ref{sec:general}. A possible way of dealing with this problem is to use a tip-tilt mirror that can change the direction of the momentum transfer associated with each laser pulse and compensate the effect of rotations, so that $\delta \boldsymbol{\mathcal{R}}$ vanishes to first order in $\Omega\, T$. This scheme was proposed in Ref.~\cite{hogan08} and has been implemented by a number of groups to overcome the effects of Earth's rotation in interferometry experiments with atomic fountains \cite{lan12,dickerson13,sugarbaker13,hauth12}.

For nonaligned gravity gradients, i.e.\ when $\mathbf{v}_\text{rec}$ does not coincide with a principal direction of the gravity gradient tensor $\Gamma (t)$,
Eqs.~\eqref{14a}-\eqref{14b} imply that
$\delta \boldsymbol{\mathcal{R}} \nparallel \mathbf{v}_\text{rec}$ and $\delta \boldsymbol{\mathcal{P}} \nparallel \mathbf{v}_\text{rec}$.
Fortunately, the mitigation strategy presented in the previous sections can be extended fairly easily to this new situation. The basic idea is to use the same kind of tip-tilt mirror scheme described above to generate an additional position displacement along the transverse directions (perpendicular to $\mathbf{v}_\text{rec}$) so that, together with the displacements generated by rotations and gravity gradients, the transverse projection of condition~\eqref{eq:mitigation1} is fulfilled:
\begin{equation}
\delta \boldsymbol{\mathcal{R}}^\perp
- (t-t_0) \frac{\delta \boldsymbol{\mathcal{P}}^\perp}{m} = 0 .
\end{equation}
On the other hand, analogously to what was done for aligned gravity gradients, one can change slightly the timing of the last beam-splitter pulse to guarantee that the projection of Eq.~\eqref{eq:mitigation1} along the longitudinal direction,
\begin{equation}
\delta \boldsymbol{\mathcal{R}}^\parallel
- (t-t_0) \frac{\delta \boldsymbol{\mathcal{P}}^\parallel}{m} = 0 ,
\end{equation}
is also satisfied.
\comment{(In this case, in addition to those due to the gravity gradient and the pulse timing asymmetry, one may also need to include possible contributions to longitudinal displacements due to higher-order terms generated by rotations and the tip-tilt mirror.)}
Finally, it should be noted that although illustrated here with condition~\eqref{eq:mitigation1}, the strategy can be straightforwardly applied to more general situations corresponding to conditions~\eqref{26} and \eqref{eq:BEC_mitigation} as well.

\section{Conclusions}
\label{sec:conclusions}

We have analyzed in detail the reduction of the integrated contrast in open atom interferometers, with particular emphasis on the case of gravity gradients, and presented a simple mitigation strategy.
\Highlight{(Earlier theoretical work on the effect of gravity gradients in atom interferometry had focused almost entirely on the calculation of the phase shift \cite{audretsch94,wolf99,peters01,borde01,borde02}.)}
For aligned gravity gradients the mitigation strategy merely involves a slight change in the timing of the last beam-splitter pulse, whereas in the nonaligned case it can be combined with schemes based on the use of tip-tilt mirrors which have been employed so far to deal with transverse relative displacements caused by rotations.
As we have shown, the strategy is very effective and in principle it enables extending the half-interferometer time beyond $T = 20\, \text{s}$ with hardly any loss of contrast due to gravity gradients. Its basic limitation is the decreasing overlap between the envelopes of the two wave packets in the superposition, which eventually becomes important for sufficiently long interferometer times. Nevertheless, this should not be a major concern in the short or mid-term future since even for very narrow momentum distributions (leading to a smaller size of the envelope at long times) with $\sigma_v \sim 10^{-3}\, v_\text{rec}$, which corresponds to $T_\text{eff} \sim 1\, \text{pK}$ for $^{87} \text{Rb}$, the effect due to a decreasing overlap of the envelopes is still very small for half-interferometer times of $T =10\, \text{s}$.

An important advantage of our proposed strategy is its ease of implementation. For aligned gravity gradients it only requires a control of the pulse timing which is well within the capabilities of current experimental set-ups. (Moreover, if that were not the case, the imperfections in controlling the pulse timing would lead themselves to a comparable loss of contrast anyway.) On the other hand, for nonaligned gravity gradients one needs to use a tip-tilt mirror scheme in order to deal with the transverse displacements, but the required degree of control has already been demonstrated in experiments that employ such a scheme for compensation of rotations (and any limitations in this respect would result, if the set-up is also being used to compensate rotational effects, in a loss of contrast due to gravity gradients comparable to that caused anyway by the remaining non-compensated rotations).
Thus, we see that the implementation is indeed much simpler (with essentially no new requirements) than for possible alternative approaches such as trying to engineer the initial state while keeping all the desirable features for long-time and high-precision interferometry.
Furthermore, the relative change in the pulse timing given by Eq.~\eqref{eq:timing}, which is of order $\| \Gamma \| \, T^2$, gives rise to additional terms in the expression for the phase shift $\delta \phi$ of the same kind and magnitude as terms already present for $\delta T = 0$. Hence, these contributions should be small enough or be taken care of by any mitigation techniques already in place, so that the required $\delta T$ does not lead to any significant reduction in sensitivity nor accuracy.

In order to apply our mitigation strategy, one needs of course to know the gravity gradient beforehand. This can be measured by other means, determined from geodetic calculations and simulations of the mass distribution near the experimental set-up or even from a calibration of $\delta T$ which eliminates the fringe pattern in the density profile of the exit ports (provided that the gravity gradient does not change much over time or it does so slowly enough). Fortunately, one does not need to know the gravity gradient very precisely: even with a 10\% inaccuracy (which is a rather conservative assumption) the method can effectively deal with the loss of contrast due to gravity gradients
\comment{up to half-interferometer times close to $T=10\, \text{s}$, as shown in Fig.~\ref{fig:gaussian_pure}.}
At this point it should be mentioned that by considering \comment{multiloop}
(e.g. with \comment{three} intermediate $\pi$ pulses) and an appropriate choice of the time between the different pulses (independent of $\Gamma$) one can achieve a vanishing final displacement $\delta \boldsymbol{\chi}$ \comment{up to a certain order} in $\| \Gamma \| \, T^2$ (and $\Omega\, T$) for time-independent (or very slowly varying) gravity gradients \cite{marzlin96}. However, for such configurations the lowest-order contributions to the phase shift from time-independent $\mathbf{g}$ (and $\Gamma$) cancel out too. These are the relevant ones for accelerometers and gravimeters as well as for tests of the weak equivalence principle (the universality of free fall) and gradiometers, where one performs differential acceleration measurements for different species or at different spatial locations. Hence, although such configurations exhibit no significant loss of contrast due to (time-independent) gravity gradients, the typical signals of interest are severely suppressed as well and they end up being of little use for those kinds of applications.

Our results are valid for atom interferometers based on Raman or Bragg pulses, and they can be straightforwardly extended to 
double diffraction schemes \cite{leveque09,giese13}: in most of the results one simply needs to double the effective momentum transfer associated with each laser pulse.
In addition, they can be applied to MZIs with several intermediate $\pi$ pulses and easily generalized to configurations with more than two beam-splitter pulses (such as Ramsey-Bord\'e interferometers) as long as only two wave packets contribute to the superposition at each exit port and these are completely distinguishable (thanks to good spatial separation or internal-state labeling).
Furthermore, as a side result we have also provided a remarkably compact derivation of the general expression for the phase shift in Appendix~\ref{sec:rep_free}, where a representation-free \cite{schleich13,kleinert14a} description of the state evolution in a light-pulse atom interferometer has been derived. This has been obtained for quadratic Hamiltonians including time-dependent uniform forces (or accelerations) and gravity gradients, characterized by $\mathbf{g}(t)$ and $\Gamma(t)$. However, it can be easily generalized to anharmonic potentials which can be locally approximated (within regions of the wave-packet size) by quadratic ones \cite{zeller14a}.
Moreover, although the effects of rotations have been described in a nonrotating frame \cite{kleinert14a}, the Hamiltonian in a rotating frame would still be quadratic and the results in Appendix~\ref{sec:rep_free} would also be applicable.

We close this section by discussing a recent proposal made in Refs.~\cite{dickerson13,sugarbaker13} to extract useful information for precision interferometry directly from the features of the fringe pattern in the density profile at the exit ports. The basic idea is that by fitting the observed density profile with a formula like Eq.~\eqref{eq:prob_density2} and an appropriate envelope (typically a Gaussian), one can in principle extract information on the displacement $\delta \boldsymbol{\chi}$ and the phase shift $\delta \Phi''$ even for a single shot \cite{miller05,muentinga13,sugarbaker13,zeller14b}. In more realistic situations various factors will lead to a reduction of the contrast in the fringe pattern, which will be better described by the following formula for the density profile: $\big|\psi_\text{I} (\mathbf{x},t) \big|^2 = (1/2) \big[ 1 + C_\text{f} \cos (\mathbf{k}_\text{fr} \cdot \mathbf{x} + \delta \Phi'') \big] \, \big|\psi_\text{c} (\mathbf{x},t) \big|^2$, with $C_\text{f} \leq 1$. Interestingly, by fitting the density profile, one can estimate the phase shift and the fringe contrast $C_\text{f}$ for every single shot, which is not possible if one considers instead the integrated atom number at each exit port. As emphasized in Refs.~\cite{dickerson13,sugarbaker13}, this can be particularly advantageous for portable inertial sensors in dynamical platforms, which may experience significant contrast fluctuations. The approach has been demonstrated with a measurement of Earth's rotation rate with a precision of $200\, \text{nrad/s}$ \cite{dickerson13} and a gyrocompass with $10\, \text{mdeg}$ precision \cite{sugarbaker13}. In both sets of experiments the measured quantities were derived from the observed fringe spacings;
moreover, in the latter controlled changes of the final displacement were introduced to modify the fringe spacing and facilitate its read-out.
Since it allows extracting information on the phase shift for nonvanishing displacements, fitting the fringe pattern of the density profiles also offers an alternative approach in those situations where gravity gradients induce a significant displacement.
A careful analysis taking into account the specific details of each particular experimental implementation may be necessary in order to determine what is the preferable method for measuring the phase shift in each case. For instance, the simpler data analysis involved in measuring the integrated particle number and the fact that is a well-proven technique which has been employed in numerous high-precision atom-interferometry experiments to date are desirable features for a space mission (and this was the choice made in the design of STE-QUEST).
On the other hand, the additional information contained in the density profile offers the possibility of a more thorough and continuous monitoring of (possibly unexpected) sources of noise and systematics.

The experiments reported in 
\highlight{Ref.~\cite{sugarbaker13}
make use of cold thermal atoms} and can be described in terms of the mixed Gaussian states studied in Sec.~\ref{sec:gaussian_mixed}. In order to discuss the point-source-interferometry (PSI) approximation considered in those references, it is convenient to decompose the initial Gaussian thermal state with $\Sigma_{xp} = 0$ as an ensemble of pure Gaussian states with $\Sigma_{xp} = 0$ and the same $\Sigma_{xx}$ as the thermal state but a smaller momentum width $\Sigma_{pp}$. These pure states have different central momenta weighed according to a Gaussian distribution, but are otherwise identical. 
As we plan to discuss in more detail in a separate publication, the treatment of
\highlight{Ref.~\cite{sugarbaker13}}
naturally corresponds to the regime where the density profile at the exit ports for each member of the ensemble exhibits no fringe pattern (although a fringe pattern may arise for the whole ensemble due to the dephasing between the different members).
In general, however, the PSI approximation has a number of limitations and can miss relevant effects associated with $\delta \boldsymbol{\mathcal{P}}_0$ which can, in contrast, be fully accounted for with the formalism that we have presented here \cite{roura14b}.

\begin{acknowledgments}
It is a pleasure to thank Stephan Kleinert and Holger Ahlers for useful and stimulating discussions as well as Naceur Gaaloul, Christian Schubert, Achim Peters and Ernst Rasel for interesting conversations and remarks.
A.~R.\ and W.~Z.\ were supported by the German Space Agency (DLR) with funds provided by the Federal Ministry of Economics and Technology (BMWi) under Grant No.~50WM1136 (QUANTUS III).
W.~P.~S.\ gratefully acknowledges the support by a Texas A\&M University Institute of Advanced Study Faculty Fellowship.
\end{acknowledgments}


\appendix

\appsection{Wave-packet trajectories in the presence of gravity gradients}
\label{sec:trajectories}

The phase-space vectors $\boldsymbol{\chi} (t)$ corresponding to the trajectories of the center of the wave packets are solutions of the classical equations of motion associated with the Hamiltonian~\eqref{eq:hamiltonian}, which can be written as
\begin{equation}
\dot{\boldsymbol{\chi}} (t) - \mathcal{H}(t) \, \boldsymbol{\chi} (t)
= \boldsymbol{\mathcal{G}}(t) + \boldsymbol{\mathcal{F}}_\text{lp} (t)
\label{A1},
\end{equation}
with
\begin{align}
\mathcal{H}(t) &=
\left( \begin{array}{cc} 0 & (1/m) \, \mathbb{1} \\ m \, \Gamma(t) & 0 \end{array} \right) 
\label{A2}, \\
\boldsymbol{\mathcal{G}}(t) &=
\left( \begin{array}{c} \mathbf{0} \\ m \mathbf{g}(t) \end{array} \right)  \label{A3},\\
\boldsymbol{\mathcal{F}}_\text{lp} (t) &= \sum_{i=1}^n  \varepsilon_i \, \delta(t-t_i)
\left( \begin{array}{c} \mathbf{0} \\ \hbar \mathbf{k}_i \end{array} \right)
\equiv \sum_{i=1}^n \delta(t-t_i) \, \boldsymbol{\mathcal{F}}_i 
\label{A4},
\end{align}
where $\boldsymbol{\mathcal{F}}_\text{lp} (t)$ accounts for the kicks from the $n$ laser pulses with $\varepsilon_i = 0, \pm1$ for each pulse and taking different values on each interferometer branch (and the corresponding classical trajectories depicted in Fig.~\ref{fig:interferometer1}).
The solutions of Eq.~\eqref{A1} are uniquely specified by the initial conditions $\boldsymbol{\chi} (t_0) = \boldsymbol{\chi}_0$ and are given by
\begin{equation}
\boldsymbol{\chi} (t) = \mathcal{T} (t,t_0)\, \boldsymbol{\chi}_0
+ (\mathcal{T}_\text{ret} \cdot \boldsymbol{\mathcal{G}}) (t)
+ (\mathcal{T}_\text{ret} \cdot \boldsymbol{\mathcal{F}}_\text{lp}) (t)
\label{A5},
\end{equation}
where the transition matrix $\mathcal{T} (t,t_0)$ satisfies the homogeneous part of Eq.~\eqref{A1} with initial condition $\mathcal{T} (t_0,t_0) = \mathbb{1}$, and we have employed the retarded propagator defined by $\mathcal{T}_\text{ret} (t,t') = \mathcal{T} (t,t') \, \theta(t-t')$ and introduced the notation
\begin{equation}
(\mathcal{T}_\text{ret} \cdot \boldsymbol{\mathcal{A}}) (t) \equiv \int^t_{t_0} dt' \, \mathcal{T}_\text{ret} (t,t')\, \boldsymbol{\mathcal{A}} (t')
\label{eq:convolution} .
\end{equation}
Moreover, one usually defines $\mathcal{T} (t_1,t_2) \equiv \mathcal{T}^{-1} (t_2,t_1)$ for $t_1 < t_2$.
Since $\mathcal{T} (t,t')$ is the transition matrix associated with the equations of motion derived from a quadratic Hamiltonian, it is a symplectic matrix, i.e.\ it leaves the symplectic form $J$ invariant \cite{abraham}:
\begin{equation}
\mathcal{T}^\mathrm{T}(t,t')\, J\, \mathcal{T}(t,t') = J
\label{eq:symplectic} .
\end{equation}

For a time-independent gravity gradient tensor the transition matrix can be straightforwardly obtained by exponentiating the matrix $\mathcal{H}(t)$:
\begin{equation}
\mathcal{T} (t,t') =
\left( \begin{array}{cc}
\cosh \left[\gamma (t-t')\right]
& \frac{1}{m\gamma} \sinh \left[\gamma (t-t')\right] \\
m \gamma\, \sinh \left[\gamma (t-t')\right]
& \cosh \left[\gamma (t-t')\right]
\end{array} \right)
\label{A6},
\end{equation}
where we have introduced $\gamma \equiv \sqrt{\Gamma}$. In order to calculate explicitly the transition matrix in Eq.~\eqref{A6} one needs to diagonalize the symmetric tensor $\Gamma$, which is always possible with an orthogonal transformation (a rotation of the coordinate axes). In this new coordinate system the motion along each one of the three axes (known as principal axes) decouples and the dynamics reduces to that of three independent one-dimensional systems. In general, some of the eigenvalues of $\Gamma$ will be negative since its trace vanishes outside gravitational sources. For those directions with negative eigenvalues one needs to write $\gamma_{jj} = i\, \omega_{jj}$ in Eq.~\eqref{A6} and the dynamics along that direction corresponds then to that of a one-dimensional harmonic oscillator, whereas for positive eigenvalues one can directly apply Eq.~\eqref{A6}, which describes the dynamics of an inverted harmonic oscillator.
In typical cases of interest in atom interferometry the condition $|\Gamma_{jj}| T^2 \ll 1$ is amply satisfied for the three principal axes and it is an excellent approximation to consider the following perturbative expansion of Eq.~\eqref{A6} up to linear order in $\Gamma$ (the expansion involves only even powers of $\gamma)$:
\begin{equation}
\mathcal{T} (t,t') \approx
\left( \begin{array}{cc}
\mathbb{1} + \frac{\Gamma}{2} (t-t')^2
& \frac{(t-t')}{m} \Big[\mathbb{1} + \frac{\Gamma}{6} (t-t')^2 \Big] \\
m \, \Gamma \, (t-t')
& \mathbb{1} + \frac{\Gamma}{2} (t-t')^2
\end{array} \right)
\label{A6b},
\end{equation}
where we neglected \comment{terms of higher order in $\Gamma (t-t')^2$.}
A similar expression (but involving one or two time integrals) can be obtained for a general time-dependent $\Gamma (t)$ by solving perturbatively Eq.~\eqref{A1}.

Note that the first two terms on the right-hand side of Eq.~\eqref{A5} are common for the two interferometer branches%
\footnote{
\comment{This does not hold in the case of branch-dependent forces discussed at the end of Appendix~\ref{sec:rep_free} or if one considers an anharmonic potential which can only be locally approximated by harmonic potentials.}
},
so that only the third term contributes to the relative displacement between the classical trajectories for the center of the wave packets in Fig.~\ref{fig:interferometer1}:
\begin{equation}
\delta \boldsymbol{\chi} (t) =
(\mathcal{T}_\text{ret} \cdot \delta\boldsymbol{\mathcal{F}}_\text{lp}) (t)
\label{A7},
\end{equation}
where $\delta\boldsymbol{\mathcal{F}}_\text{lp}$ simply corresponds to the difference between the laser kicks, as given by Eq.~\eqref{A4}, for the two trajectories.
This also implies that the evolution of the displacement after the last beam splitter takes the following simple form:
\begin{equation}
\delta \boldsymbol{\chi} (t) = \mathcal{T}(t,t_\text{bs}) \, \delta \boldsymbol{\chi} (t_\text{bs})
\label{eq:displacement_bs} ,
\end{equation}
where we used the definition of $\mathcal{T}_\text{ret}$ in terms of the transition matrix and the composition of transition matrices.

Making use of Eq.~\eqref{A6b}, one obtains from Eq.~\eqref{A7} the following displacement for a MZI, where the effective momentum transfer $\hbar \mathbf{k}$ is the same for all the laser pulses and induces a recoil velocity $\mathbf{v}_\text{rec} = \hbar \mathbf{k} / m$:
\begin{align}
\delta \boldsymbol{\mathcal{R}}(t_\text{bs}) &\approx - \mathbf{v}_\text{rec} \, \delta T
+ \left(\Gamma\, T^2 \right) \mathbf{v}_\text{rec} \, T
\left[ 1 + \frac{\delta T}{T} - \frac{1}{6} \left( \frac{\delta T}{T} \right)^3 \right]
\label{eq:displacement_mzi_r} , \\
\delta \boldsymbol{\mathcal{P}}(t_\text{bs}) & \approx
\left(\Gamma\, T^2 \right) m \mathbf{v}_\text{rec} \left[1 - \frac{1}{2} \left( \frac{\delta T}{T} \right)^2 \right]
\label{eq:displacement_mzi_p} ,
\end{align}
with time separations between the central $\pi$ pulse and the initial and final $\pi/2$ pulses of $T$ and $T + \delta T$ respectively.

\appsection{Representation-free description of the state evolution in an atom interferometer}
\label{sec:rep_free}

In this appendix we will derive a useful representation-free description for the state evolution of the atoms in an atom interferometer \comment{and establish the connection with similar approaches in the literature.}

Let us consider an initial state
\begin{equation}
|\psi (t_0)\rangle =
\hat{\mathcal{D}}(\boldsymbol{\chi}_0)\,  |\psi_\text{c} (t_0) \rangle
\label{B1},
\end{equation}
where $\hat{\mathcal{D}}(\boldsymbol{\chi})$ is the displacement operator defined in Eq.~\eqref{eq:displacement}, applied here with an initial displacement $\boldsymbol{\chi}_0$, and $|\psi_\text{c} (t_0)\rangle$ will be referred to as the \emph{centered wave packet}. 
Although the results below hold independently of this, it is natural to choose $\boldsymbol{\chi}_0 = \langle\psi (t_0)| \, \hat{\boldsymbol{\xi}} \, |\psi (t_0)\rangle$ so that the centered state $|\psi_\text{c} (t_0)\rangle$ is uniquely specified.
We will focus on the evolution generated by the quadratic Hamiltonian of Eq.~\eqref{eq:hamiltonian}. It is convenient to introduce its decomposition
\begin{equation} 
\hat{H} = \hat{H}_0 + \hat{V}_g + V_0
\label{B2} ,
\end{equation}
in terms of the purely quadratic and linear parts
\begin{align}
\hat{H}_0 (t) &= \frac{1}{2m} \hat{\mathbf{p}}^\text{T} \hat{\mathbf{p}}
- \frac{m}{2} \hat{\mathbf{x}}^\text{T} \Gamma (t)\, \hat{\mathbf{x}} \label{B3}, \\
\hat{V}_g (t) &= - m\, \mathbf{g}^\text{T} (t) \, \hat{\mathbf{x}} 
\label{B4}.
\end{align}
\highlight{The spatially independent potential $V_0(t)$, which does not alter the classical equations of motion,
is proportional to the identity operator as far as the external degrees of freedom are concerned. It simply generates a time-dependent phase, but can give rise to nontrivial effects if it takes different values along each interferometer branch.
In particular, it can naturally account for possible changes in the the rest mass $m c^2$ (associated with different internal energies as the internal state changes along each branch) as well as other effects mentioned below in Appendix~\ref{sec:phase_shift}.}

\subsection{Purely quadratic part}

With respect to the Hamiltonian $\hat{H}_0$ the phase-space vector $\hat{\boldsymbol{\xi}} = (\hat{\mathbf{x}},\hat{\mathbf{p}})^\text{T}$, consisting of the position and momentum operators, exhibits the following relationship between the Heisenberg and the Schr\"odinger pictures: $\hat{\boldsymbol{\xi}}_\text{H} = \hat{U}^\dagger_0 (t,t_0)\,\hat{\boldsymbol{\xi}} \,\hat{U}_0 (t,t_0) = \mathcal{T}(t,t_0) \, \hat{\boldsymbol{\xi}}$,
where $\mathcal{T}(t,t_0)$ coincides with the transition matrix associated to the homogeneous part of Eq.~\eqref{A1}, which is a consequence of linearity and the fact that the Heisenberg equations have the same form as the classical equations of motion.
Such a relation can be inverted to give $\hat{U}_0 (t,t_0)\,\hat{\boldsymbol{\xi}} \,\hat{U}^\dagger_0 (t,t_0) = \mathcal{T}^{-1}(t,t_0) \, \hat{\boldsymbol{\xi}}$. This together with Eq.~\eqref{eq:symplectic} implies the following result for the displacement operator:
\begin{equation}
\hat{U}_0 (t,t_0)\,\hat{\mathcal{D}} \big( \boldsymbol{\chi}_0 \big)\, \hat{U}^\dagger_0 (t,t_0)
= \hat{\mathcal{D}} \big( \boldsymbol{\chi}_\text{h}(t) \big)
\label{B5},
\end{equation}
where $\boldsymbol{\chi}_\text{h}(t) = \mathcal{T}(t,t_0) \, \boldsymbol{\chi}_0$ corresponds to the classical phase-space trajectory associated with the Hamiltonian $H_0$ and with initial condition $\boldsymbol{\chi}_\text{h} (t_0) = \boldsymbol{\chi}_0$.

Taking all that into account and inserting the resolution of the identity $I = \hat{U}^\dagger_0 (t,t_0) \,\hat{U}_0 (t,t_0)$ after the displacement operator on the right-hand side of Eq.~\eqref{B1}, the evolution of that state generated by the purely quadratic part of the Hamiltonian can be written in the following suggestive way:
\begin{equation}
|\psi (t)\rangle = \hat{U}_0 (t,t_0) \, |\psi (t_0)\rangle
= \hat{\mathcal{D}} \big( \boldsymbol{\chi}_\text{h} (t) \big) \, |\psi_\text{c} (t) \rangle
\label{B6},
\end{equation}
with $|\psi_\text{c} (t) \rangle = \hat{U}_0 (t,t_0) \, |\psi_\text{c} (t_0) \rangle$ and
\begin{equation}
\boldsymbol{\chi}_\text{h} (t) = \mathcal{T}(t,t_0) \, \boldsymbol{\chi}_0 .
\end{equation}

\subsection{Purely linear part and laser pulses}

Let us consider the effect of the \emph{laser pulses} first. \comment{Focusing on the external degrees of freedom (the internal state can be easily taken into account by appropriately labeling the segments of the trajectories after each pulse and considering only the interference between wave packets with the same internal state at the exit ports),} the effect of idealized beam-splitter and mirror pulses can be characterized (for the $n$-th pulse) by a phase factor $e^{i \varepsilon_n\varphi_n}\, \highlight{e^{-i |\varepsilon_n| \frac{\pi}{2}}}$ times a displacement operator $e^{i \varepsilon_n \mathbf{k}_n^\text{T} \hat{\mathbf{x}}} = \hat{\mathcal{D}} \big( \boldsymbol{\mathcal{F}}_n \big)$ acting on every wave packet. The integer $\varepsilon_n$, the laser phase $\varphi_n$, the momentum transfer $\hbar\mathbf{k}_n$ and the associated phase-space displacement $\boldsymbol{\mathcal{F}}_n$ [defined in Eq.~\eqref{A4}] depend on both the pulse number $n$ and the interferometer branch (the classical trajectory assigned to each wave packet). In addition, beam-splitter pulses generate a superposition corresponding to two of such operators with different values of $\varepsilon_n$ acting on the same wave packet.
Given a state of the form $\hat{\mathcal{D}} \big( \boldsymbol{\chi} (t_n^-) \big) \, |\psi_\text{c} (t_n) \rangle$ right before, the action of a laser pulse at time $t_n$ generates the following state:
\begin{align}
|\psi (t_n)\rangle &=
\highlight{e^{i (\varepsilon_n\varphi_n - |\varepsilon_n| \frac{\pi}{2})}}
\, \hat{\mathcal{D}} \big( \boldsymbol{\mathcal{F}}_n \big) \,
\hat{\mathcal{D}} \big( \boldsymbol{\chi} (t_n^-) \big) \, |\psi_\text{c} (t_n) \rangle
\nonumber \\
&= \highlight{e^{i (\varepsilon_n\varphi_n - |\varepsilon_n| \frac{\pi}{2})}}
\, e^{-\frac{i}{2 \hbar} \boldsymbol{\mathcal{F}}_n^\text{T} J\,
\boldsymbol{\chi} (t_n^-)} \, \nonumber \\
& \quad \quad \times \hat{\mathcal{D}} \big( \boldsymbol{\chi} (t_n^-)
+ \boldsymbol{\mathcal{F}}_n  \big) \,
| \psi_\text{c} (t_n) \rangle
\label{eq:laser_kick1},
\end{align}
where we used the composition formula in Eq.~\eqref{5}.
Evolution between pulses proceeds as described in the previous subsection, so that the state at some time $t > t_n$ before the action of any further pulse is given by
\begin{equation}
|\psi (t)\rangle = e^{i \Phi(t)}\, \hat{\mathcal{D}} \big( \boldsymbol{\chi}(t)  \big) \, | \psi_\text{c} (t) \rangle
\label{eq:laser_kick2},
\end{equation}
where the displacement $\boldsymbol{\chi}(t)$ corresponds to the classical trajectory with initial conditions $\boldsymbol{\chi}(t_0) = \boldsymbol{\chi}_0$ and including the kicks from the laser pulses:
\begin{align}
\boldsymbol{\chi}(t) & = \mathcal{T}(t,t_n) \, \boldsymbol{\chi}(t_n^-)
+ \mathcal{T}(t,t_n)\, \boldsymbol{\mathcal{F}}_n\,   \nonumber \\
& = \boldsymbol{\chi}_\text{h} (t)
+ \big( \mathcal{T}_\text{ret} \cdot \boldsymbol{\mathcal{F}}_\text{lp} \big) (t)
\label{eq:laser_displacement} ,
\end{align}
with $\boldsymbol{\mathcal{F}}_\text{lp} $ defined in Eq.~\eqref{A4}. The phase $\Phi(t)$ in Eq.~\eqref{eq:laser_kick2} is in turn given by
\begin{equation}
\Phi(t) = \varphi - \frac{1}{2\hbar} \sum_{j=1}^n \boldsymbol{\mathcal{F}}_j^\text{T} J\, \boldsymbol{\chi} (t_j)
\label{eq:laser_phase},
\end{equation}
with $\varphi = \sum_{j=1}^n \big( \varepsilon_j\, \varphi_j - \highlight{|\varepsilon_j| \, \pi/2} \big)$.
In Eqs.~\eqref{eq:laser_displacement}-\eqref{eq:laser_phase} we have already provided the general result for $n$ pulses. It can be easily proved by induction. First, one uses Eq.~\eqref{eq:laser_kick1} to prove the result for $n=1$. Next, assuming that the result holds for $n-1$ pulses, one can use Eq.~\eqref{eq:laser_kick1} again to show that it then holds for $n$ pulses too.

Let us now turn our attention to the \emph{purely linear} term in the Hamiltonian. The quickest way of generalizing the previous results in order to include the effect of the linear term is by regarding it as equivalent to the continuum limit $\delta t \to 0$ of $r$ pulses with $\boldsymbol{\mathcal{F}}_j = \delta t \, \boldsymbol{\mathcal{G}}(t_j)$, where $ \boldsymbol{\mathcal{G}}(t)$ is defined in Eq.~\eqref{A3}, $t_j = t_0 + j\, \delta t$ with $j=1,\ldots,r$ and $r\, \delta t = t-t_0$.
Eq.~\eqref{eq:laser_displacement} is then generalized to Eq.~\eqref{A5}, and Eq.~\eqref{eq:laser_phase} becomes
\begin{equation}
\Phi(t) = \varphi - \frac{1}{2\hbar} \int_{t_0}^t \! dt' \Bigg( \Big[ \boldsymbol{\mathcal{F}}_\text{lp}^\text{T}(t') + \boldsymbol{\mathcal{G}}^\text{T}(t') \Big] J\, \boldsymbol{\chi} (t') \,+\, 2V_0 (t') \Bigg)
\label{eq:laser_phase2},
\end{equation}
\highlight{where we have also included the contribution from the \emph{spatially independent} potential $V_0(t)$.}

\subsection{Phase shift}
\label{sec:phase_shift}

At the exit ports one has a superposition of several wave packets, which can each be evolved as described in the previous subsection. For interferometers where two wave packets contribute to each exit port, as depicted for instance in Fig.~\ref{fig:interferometer1}, the superposition is given by Eq.~\eqref{4}. The displacement $\delta\boldsymbol{\chi}$ can be calculated as explained in Appendix~\ref{sec:trajectories}. Here we present a compact but complete derivation of the result for the phase shift $\delta\Phi$.

First one takes the expression for $\delta\Phi$ provided after Eq.~\eqref{4} and substitutes $\Phi_1$ and $\Phi_2$ by the result in Eq.~\eqref{eq:laser_phase2}. (For the moment we do not consider the contributions from the purely linear term of the potential and take $\boldsymbol{\mathcal{G}}=0$ so that the expressions are shorter, but they can be easily added at the end, as done in the previous subsection.)
Next, one rewrites the result in terms of the semisum and difference variables $\boldsymbol{\bar{\chi}} = (\boldsymbol{\chi}_1 + \boldsymbol{\chi}_2)/2$ and $\delta\boldsymbol{\chi} = \boldsymbol{\chi}_2 - \boldsymbol{\chi}_1$. One also proceeds in the same way with the sets of pulse displacements $\boldsymbol{\mathcal{F}}_j^{(1)}$ and $\boldsymbol{\mathcal{F}}_j^{(2)}$ associated with the two classical trajectories as well as the phases $\varphi_1$ and $\varphi_2$. Together with several additional steps described in more detail below, this leads to the following intermediate and final result:
\begin{widetext}

\begin{align}
\delta\Phi(t)
& = \Phi_2 - \Phi_1 + \frac{1}{2\hbar}\, \boldsymbol{\chi}_1^\text{T} J \, \boldsymbol{\chi}_2
= \delta \varphi - \frac{1}{2\hbar}  \int_{t_0}^t dt'
\Big[ \delta \boldsymbol{\mathcal{F}}_\text{lp}^\text{T}(t')\, J\, \boldsymbol{\bar{\chi}} (t')
+ \boldsymbol{\bar{\mathcal{F}}}_\text{lp}^\text{T}(t')\, J\, \delta \boldsymbol{\chi} (t') + 2\, \delta V_0 (t') \Big]
- \frac{1}{2\hbar}\, \delta \boldsymbol{\chi}^\text{T}(t)\, J \, \boldsymbol{\bar{\chi}}(t)
\nonumber \\
& = \delta \varphi - \frac{1}{2\hbar}  \int_{t_0}^t dt'
\Big[ \delta \boldsymbol{\mathcal{F}}_\text{lp}^\text{T}(t')\, J\, \boldsymbol{\bar{\chi}} (t')
- \delta \boldsymbol{\chi}^\text{T} (t')\, J\, \boldsymbol{\bar{\mathcal{F}}}_\text{lp}(t')
+ \delta \dot{\boldsymbol{\chi}}^\text{T}(t')\, J \, \boldsymbol{\bar{\chi}}(t')
+ \delta \boldsymbol{\chi}^\text{T}(t')\, J \, \boldsymbol{\dot{\bar{\chi}}}(t')  + 2\, \delta V_0 (t') \Big]
\nonumber \\
& = \delta \varphi - \frac{1}{\hbar} \int_{t_0}^t dt'
\Big[ \delta \boldsymbol{\mathcal{F}}_\text{lp}^\text{T}(t')\, J\, \boldsymbol{\bar{\chi}} (t')
+ \delta V_0 (t') \Big]
\label{eq:phase_shift} .
\end{align}
\end{widetext}
In writing the second line, we took the transpose of the second term of the integrand (it is always possible since it is a scalar), which led to a change of sign because $J^\mathrm{T} = - J$. We also wrote the last term on the first line as the time integral from $t_0$ to $t$ of its time derivative, taking into account that $\delta \boldsymbol{\chi} (t_0) = 0$ for $t_0$ before the first beam splitter.
Finally, the following steps were carried out in the last equality. First, the time derivatives were substituted using the relations $\delta \dot{\boldsymbol{\chi}} = \mathcal{H} \, \delta\boldsymbol{\chi} + \delta \boldsymbol{\mathcal{F}}_\text{lp}$ and $\boldsymbol{\dot{\bar{\chi}}} = \mathcal{H} \, \boldsymbol{\bar{\chi}} + \boldsymbol{\mathcal{\bar{F}}}_\text{lp}$ that follow from Eq.~\eqref{A1}.
Second, we used the identity $\mathcal{H}^\text{T} J = -J\, \mathcal{H}$, which holds for any quadratic Hamiltonian system. Of the two remaining terms one cancels the second term in the integrand, whereas the other is identical to the first term in the integrand, which gets doubled.

We can now include the contributions from the purely linear term of the Hamiltonian by proceeding as done at the end of the previous subsection. All that one needs to do is to use the full solution in Eq.~\eqref{A5}, including $\boldsymbol{\mathcal{G}}$, with $\boldsymbol{\bar{\mathcal{F}}}_\text{lp}$ in place of $\boldsymbol{\mathcal{F}}_\text{lp}$.
Eq.~\eqref{eq:phase_shift} agrees then with the general result obtained by Antoine and Bord\'e in Refs.~\cite{antoine03a,antoine03b} (see also Refs.~\cite{audretsch94,wolf99,peters01,borde01,borde02} for earlier results including the effects of gravity gradients).
Furthermore, one can easily generalize the result to the case with different values of $\boldsymbol{\mathcal{G}}$ for each branch.
This can be relevant for instance when there is a uniform force whose value depends on the internal state of the atoms acting in an interferometer with different internal states in each branch.
\highlight{[It is caused, for example, by time-dependent magnetic field gradients in Raman-based interferometers due the quadratic Zeeman effect; furthermore, time-dependent homogeneous magnetic fields give rise in that case to nontrivial effects through $\delta V_0(t')$.]}
Considering branch-dependent $\boldsymbol{\mathcal{G}}$ and $V_0$  is also useful for describing the effects of anharmonic potentials which can be locally approximated (within regions of the size of each wave packet) by quadratic ones, as discussed in detail in Ref.~\cite{zeller14a}.
The new result can be written in a compact form as
\begin{align}
\delta\Phi(t) &= \delta\varphi - \frac{1}{\hbar} \int_{t_0}^t \! dt' \, \Bigg(
\Big[ \delta\boldsymbol{\mathcal{F}}_\text{lp}^\text{T}(t') + \delta\boldsymbol{\mathcal{G}}^\text{T}(t') \Big]
J\, \boldsymbol{\bar{\chi}} (t') \nonumber \\
& \qquad \qquad \qquad \qquad \;
+ \delta V_0 (t') \Bigg)
\label{eq:phase_shift2},
\end{align}
where $\boldsymbol{\bar{\chi}} (t')$ is given by Eq.~\eqref{A5} with $\boldsymbol{\bar{\mathcal{F}}}_\text{lp}$ and $\boldsymbol{\bar{\mathcal{G}}}$.

Finally, it is sometimes convenient to write the result in an alternative way which explicitly displays all the dependence on the initial conditions. When doing so, Eq.~\eqref{eq:phase_shift2} becomes
\begin{align}
\delta\Phi(t) & = \delta\varphi - \frac{1}{\hbar} \int_{t_0}^t dt'\, \delta V_0 (t')
- \frac{1}{\hbar}\, \delta \boldsymbol{\chi}^\text{T}(t)\, J \, 
\mathcal{T}(t,t_0) \, \boldsymbol{\chi}_0
\nonumber \\
&\quad - \frac{1}{\hbar} \int_{t_0}^t dt' \int_{t_0}^{t'} dt''
\Big[ \delta\boldsymbol{\mathcal{F}}_\text{lp}^\text{T}(t') + \delta\boldsymbol{\mathcal{G}}^\text{T}(t') \Big] J\,
\nonumber\\
& \quad \quad \quad \quad \times \mathcal{T}(t',t'') \,
\Big[ \boldsymbol{\bar{\mathcal{F}}}_\text{lp}(t'') + \boldsymbol{\bar{\mathcal{G}}}(t'') \Big]
\end{align}
where we have used Eq.~\eqref{A5} for $\boldsymbol{\bar{\chi}} (t')$, the identity $J\, \mathcal{T} (t',t_0) = \mathcal{T}^\text{T} (t,t')\, J\, \mathcal{T} (t,t_0)$, which follows from Eq.~\eqref{eq:symplectic}, and the expression for $\delta \boldsymbol{\chi}^\text{T}(t)$ in terms of $\delta \boldsymbol{\mathcal{F}}_\text{lp}^\mathrm{T}$ and $\delta \boldsymbol{\mathcal{G}}^\mathrm{T}$ obtained from Eq.~\eqref{A5}.

\highlight{It should be noted that although we have focused on the Hamiltonian~\eqref{B2}, the derivation in this appendix can be directly applied to a general quadratic Hamiltonian (including general linear terms) as well. In particular, the results can be straightforwardly applied to calculations in rotating frames, where new quadratic terms associated with the Coriolis and centrifugal forces arise.}

\appsection{Late-time free evolution}
\label{sec:late-time}

It is well known that the free evolution of a wave packet in position representation takes a simple form at late times,
as we will see.
One starts by noting that the free evolution of a wave packet $| \psi(t) \rangle$ can be straightforwardly calculated in momentum representation, i.e.\ for $\tilde{\psi} (\mathbf{p},t) = \langle \mathbf{p} | \psi(t) \rangle$. The evolution in position representation can then be directly obtained by an inverse Fourier transform:
\begin{equation}
\psi (\mathbf{x},t) = 
\int \frac{d^3p}{(2\pi \hbar)^{3/2}}\,
e^{\frac{i}{\hbar} \mathbf{p}^\mathrm{T} \mathbf{x}} \,
e^{-\frac{i}{\hbar} (\mathbf{p}^2 / 2m) (t-t_0)} \, \tilde{\psi} (\mathbf{p},t_0)
\label{C1},
\end{equation}
which can be conveniently rewritten as
\begin{equation}
\psi (\mathbf{x},t) = e^{\frac{i}{\hbar} \frac{m}{2 \Delta t} \mathbf{x}^2} \!
\int \! \frac{d^3p}{(2\pi \hbar)^{3/2}} \,
e^{-\frac{i}{\hbar} \frac{\Delta t}{2m} \big(\mathbf{p} - \big(\frac{m}{\Delta t}\big) \mathbf{x} \big)^2} 
\tilde{\psi} (\mathbf{p},t_0) 
\label{C2},
\end{equation}
with $\Delta t = (t-t_0)$.
The integral can be exactly evaluated in certain cases, e.g.\ for Gaussian wave packets. Moreover, for sufficiently late times one can use a saddle-point approximation (also known as stationary-phase approximation) and obtain the following result:
\begin{equation}
\psi (\mathbf{x},t) \approx \left(\frac{m}{i\, \Delta t}\right)^\frac{3}{2}
e^{\frac{i}{\hbar} \frac{m}{2 \Delta t} \mathbf{x}^2}\,
\tilde{\psi} (m\, \mathbf{x}/\Delta t,t_0)
\label{C3}.
\end{equation}
This is a good approximation provided that the characteristic scales $\Delta \mathbf{p}$ over which $\tilde{\psi} (\mathbf{p},t_0)$ changes significantly are much larger than the width of the Gaussian peak in the saddle-point approximation, i.e.\ that $\Delta p^i \gg (m\hbar/\Delta t)^{1/2}$ for $i=1,2,3$.

It is worth mentioning that the result just derived can also be easily obtained within a phase-space formulation in terms of the Wigner function. One simply needs to take into account that the Wigner function evolves like a classical phase-space distribution, following the trajectories of a free particle in phase space, and that this leads to a shear (squeezing plus tilting) of the initial distribution. The result is equivalent to Eq.~\eqref{C3} once one takes into account the rescaling of the spatial coordinates on the right-hand side of that equation and the fact that the spatially dependent phase factor leads to a sheared Wigner function, as illustrated by Eqs.~\eqref{eq:wigner_BEC1}-\eqref{eq:wigner_BEC2} below.

\appsection{Description of an expanding BEC}
\label{sec:BEC_evolution}

For most purposes the expansion dynamics of a BEC released from a  (possibly time-dependent) trap can be satisfactorily described in terms of a mean-field approximation 
based on the Gross-Pitaevskii equation, which has the following form:
\begin{equation}
i \hbar \frac{\partial}{\partial t} \psi(\mathbf{x},t) = \left[ -\frac{\hbar ^2}{2m} \boldsymbol{\nabla} ^2 + V(\mathbf{x},t) + g |\psi(\mathbf{x},t)|^2 \right]\psi(\mathbf{x},t).
\label{eq:GP}
\end{equation}
where $m$ denotes the mass of the atoms and the coupling constant  $g$ is related to the $s$-wave scattering length $a_s$ according to $g = 4\pi\hbar^2a_s/m$.
Moreover, for (locally) quadratic potentials a separation of the dynamics of the center of the wave packet analogous to that described in Appendix~\ref{sec:rep_free} for non-interacting particles is also possible \cite{nandi07,eckart08,arnold,zeller14a} and it matches quite naturally with the non-interacting case when the BEC becomes dilute enough and the effects of the non-linearities can be neglected.
We will, therefore, focus throughout the rest of this appendix on the dynamics of a centered condensate evolving in a purely quadratic potential of the form
\begin{equation}
V(\mathbf{x},t) = \frac{m}{2} \mathbf{x}^\mathrm{T} \Omega^2(t) \mathbf{x} ,
\end{equation}
where $\Omega^2(t)$ is a symmetric and initially positive-definite matrix.

\subsection{Scaling approach}
\label{sec:scaling}

For purely quadratic potentials there is a very useful approach for extending the Thomas-Fermi approximation to time-dependent potentials. Within the hydrodynamic description of the condensate, regarded as a superfluid, it essentially amounts to considering a rescaled coordinate system locally comoving with the expanding fluid, i.e.\ where every fluid element is at rest \cite{dalfovo99}. The approach can also be implemented directly at the level of the Gross-Pitaevskii equation \cite{castin96,eckart08}.
One starts by introducing the linearly rescaled coordinates 
\begin{equation}
\boldsymbol{\zeta} = \Lambda^{-1}(t) \, \mathbf{x}
\label{eq:zeta},
\end{equation}
where $\Lambda (t)$ is a time-dependent matrix to be determined below.
Next, one considers the following rescaling of the wave function:
\begin{equation}
\psi (\mathbf{x},t) = \big(\! \det \Lambda(t) \big)^{-\frac{1}{2}} \, e^{i \beta(t)} \,
e^{i \Phi(\boldsymbol{\zeta},t)} \psi_\Lambda (\boldsymbol{\zeta},t)
\label{eq:scaling} ,
\end{equation}
where the prefactor with the square root of the determinant guarantees a simple normalization condition for the new wave function $\psi_\Lambda (\boldsymbol{\zeta},t)$ in terms of the rescaled coordinates $\boldsymbol{\zeta}$, namely $\int d^3\zeta \, \left|\psi_\Lambda (\boldsymbol{\zeta},t) \right|^2 = N$.
The spatially dependent and independent phases $\Phi(\boldsymbol{\zeta},t)$ and $\beta(t)$ are given respectively by
\begin{equation}
\beta(t) = -\frac{\mu}{\hbar} \int^t_{t_0} dt' \, \big( \det \Lambda(t') \big)^{-1}
\label{eq:beta} ,
\end{equation}
\begin{equation}
\Phi(\boldsymbol{\zeta},t) = \frac{m}{2\hbar} \, \boldsymbol{\zeta}^\mathrm{T} A(t) \, \boldsymbol{\zeta}
\label{eq:Phi} ,
\end{equation}
where
\begin{equation}
A(t) = \Lambda^\mathrm{T} \, \frac{d\Lambda}{dt} .
\end{equation}
As discussed below, with this choice the phase $\Phi(\boldsymbol{\zeta},t)$ will account for most of the kinetic energy of the expanding BEC in the Thomas-Fermi regime.

Substituting Eq.~\eqref{eq:scaling} into the original Gross-Pitaevskii equation~\eqref{eq:GP} 
and demanding that $\Lambda(t)$ satisfies the ordinary differential equation
\begin{equation}
 \Lambda^\mathrm{T}(t)\left( \frac{d^2 \Lambda}{d t^2} + \Omega^2(t) \Lambda(t) \right) = \frac{\Omega^2(t_0)}{\det \Lambda(t)}
\label{eq:lambda},
\end{equation}
with initial conditions
\begin{equation}
\Lambda(t_0)=\mathbb{1}, \qquad \dot{\Lambda}(t_0)=0
\label{eq:initial_conds},
\end{equation}
one finds that the equation governing the dynamics of $\psi_\Lambda (\boldsymbol{\zeta},t)$ takes the following simple and convenient form:
\begin{align}
i \hbar \left. \frac{\partial \psi_\Lambda}{\partial t} \right|_\zeta
= & -\frac{\hbar ^2}{2m} \boldsymbol{\nabla}_\zeta^\mathrm{T}
\big(\Lambda^\mathrm{T} \Lambda \big)^{-1}  \boldsymbol{\nabla}_\zeta \, \psi_\Lambda \nonumber \\
& + \frac{1}{\det \Lambda} \left[ \frac{m}{2} \, \boldsymbol{\zeta}^\mathrm{T} \, \Omega^2 (t_0) \, \boldsymbol{\zeta}  + g\, |\psi_\Lambda|^2 - \mu\right] \psi_\Lambda
\label{eq:GP2} .
\end{align}
If one neglects the first term on the right-hand side of Eq.~\eqref{eq:GP2}, its solution for an expanding condensate initially in the ground state of the trapping potential becomes time independent and coincides with the Thomas-Fermi approximation of the ground-state wave function:
\begin{align}
\psi_\Lambda (\boldsymbol{\zeta},t) \approx \psi_\mathrm{TF}(\boldsymbol{\zeta})
& = \frac{1}{\sqrt{g}} \Big( \mu - \boldsymbol{\zeta}^\mathrm{T} \Omega^2(t_0) \boldsymbol{\zeta} \Big)_+^\frac{1}{2}
\nonumber \\
& = \left( \frac{15}{8\pi} \frac{N}{\det \Sigma} \right)^{\frac{1}{2}} \Big( 1 - \boldsymbol{\zeta}^\mathrm{T} \Sigma^{-1} \boldsymbol{\zeta} \Big)_+^\frac{1}{2}
\label{eq:psi_lambda} ,
\end{align}
where the constant $\mu$ is determined by the total number of atoms in the condensate and the subindex $+$ has been introduced to denote that the function vanishes when the argument of the square root becomes negative. 
Having neglected the first term on the right-hand side of Eq.~\eqref{eq:GP2} corresponds to neglecting the contribution to the kinetic term due to the gradient of the envelope of the wave function, which constitutes a natural generalization of the Thomas-Fermi approximation to the time-dependent case and is equivalent to neglecting the so-called quantum pressure term in the hydrodynamic description.
In contrast with the ground state case, for an expanding BEC there is a significant contribution from the kinetic term in Eq.~\eqref{eq:GP} (one can easily understand from energy conservation that all the initial interaction energy associated with the nonlinear term is eventually transformed into kinetic energy), but most of it is accounted for by the spatially dependent phase~\eqref{eq:Phi} provided that the de Broglie wavelength associated with the characteristic momenta of the expanding superfluid is much shorter than the size of the envelope (the rescaled Thomas-Fermi radius).

It is clear that only the symmetric part of the matrix $A(t)$ contributes to Eq.~\eqref{eq:Phi}. Moreover, the conditions~\eqref{eq:lambda}-\eqref{eq:initial_conds} fulfilled by $\Lambda (t)$ guarantee that $A(t)$ is symmetric itself. This can be shown as follows. First, one notes that the initial conditions~\eqref{eq:initial_conds} imply $A(t_0) = A^\mathrm{T}(t_0) = 0$. Next, one can check that Eq.~\eqref{eq:lambda} implies $\dot{A}(t) - \dot{A}^\mathrm{T}(t) = 0$. It then follows that $A(t) = A^\mathrm{T}(t) \ \forall \, t$.
\pagebreak[3]

\comment{Next, we briefly discuss the phase-space description for an expanding BEC, which provides useful additional insight.}
The Wigner function for the approximate scaling solution can be obtained by substituting Eq.~\eqref{eq:scaling} into Eq.~\eqref{eq:wigner_def} and employing the time-independent approximation \eqref{eq:psi_lambda} for $\psi_\Lambda (\boldsymbol{\zeta},t)$.
\highlight{(A more detailed calculation can be found in Ref.~\cite{zeller14a}.)}
In the original unrescaled coordinates it is given by
\begin{widetext}
\begin{equation}
W(\mathbf{x},\mathbf{p},t)
= \frac{1}{ \det \Lambda(t)} \int \frac{d^3\Delta}{(2\pi\hbar)^3} \,
e^{i \mathbf{p}^\mathrm{T} \boldsymbol{\Delta}/ \hbar} \,
e^{- i \frac{m}{\hbar} \, \mathbf{x}^\mathrm{T} \dot{\Lambda} \Lambda^{-1}
\boldsymbol{\Delta}} \, 
\psi_\text{TF} \big( \Lambda^{-1} (\mathbf{x} + \boldsymbol{\Delta}/2) \big) \,
\psi_\text{TF} \big( \Lambda^{-1} (\mathbf{x} - \boldsymbol{\Delta}/2) \big)
\label{eq:wigner_BEC1}.
\end{equation}
Restoring the rescaled variable $\boldsymbol{\zeta}$ and introducing the analogously rescaled variable $\boldsymbol{\bar{\Delta}} = \Lambda^{-1} \boldsymbol{\Delta}$ together with the momentum $\mathbf{p}_\zeta = \Lambda^\mathrm{T} \mathbf{p}$, it becomes
\begin{equation}
\mathcal{W} ( \boldsymbol{\zeta},\mathbf{p}_\zeta,t )
= \int \frac{d^3 \bar{\Delta}}{(2\pi\hbar)^3} \,
e^{i \mathbf{p}_\zeta^\mathrm{T} \boldsymbol{\bar{\Delta}}/ \hbar} \,
e^{- i \frac{m}{\hbar} \, \boldsymbol{\zeta}^\mathrm{T} A(t) \, \boldsymbol{\bar{\Delta}}} \,
\psi_\text{TF} ( \boldsymbol{\zeta} + \boldsymbol{\bar{\Delta}}/2  ) \,
\psi_\text{TF} ( \boldsymbol{\zeta} - \boldsymbol{\bar{\Delta}}/2 ) 
= W_\text{TF} \big( \boldsymbol{\zeta},
\mathbf{p}_\zeta - m\, A(t)\, \boldsymbol{\zeta} \big)
\label{eq:wigner_BEC2}.
\end{equation}
\end{widetext}
%
This result has a very suggestive form and shows that the Wigner function at later times is given by a sheared version of the initial Wigner function for the Thomas-Fermi ground state with the shearing degree (the amount of tilting) controlled by the matrix $A(t)$, which vanishes at the initial time. This shear is directly related to the spatially dependent phase factor in the wave function, with a quadratic dependence on the position of the exponent, and it follows from the conversion of the nonlinear interaction energy into kinetic energy.
Furthermore, given the late-time behavior of $A(t)$, linear in time, and rescaling back to the original coordinates $(\mathbf{x}, \mathbf{p})$, one finds that the result is compatible with free evolution at late times and with the result of Appendix~\ref{sec:late-time}.
Finally, since most of the support of the Thomas-Fermi Wigner function $W_\text{TF} \big( \boldsymbol{\zeta},
\mathbf{p}_\zeta)$ (which also exhibits tails with smaller amplitude oscillations and negative values) corresponds to a region of spatial width $2 R_\text{TF}$ and momentum width $\Delta p \sim \hbar / R_\text{TF}$, the time evolution leads to a squeezing and tilting of this region which looks qualitatively similar to the evolution of the Gaussian states considered in Sec.~\ref{sec:gaussian}. Thus, the interpretation of the loss of contrast and the mitigation strategy in terms of the alignment of squeezed distributions in phase-space also applies to the scaling solution for expanding BECs. And the same can be said about the generic late-time behavior described in Appendix~\ref{sec:late-time}.
\highlight{In fact, one can provide a unified treatment for all these cases, as explicitly shown in Ref.~\cite{zeller14a}.}

\subsection{Momentum distribution at intermediate times}
\label{sec:momentum_distrib}

An interesting result within the scaling approximation, pointed out in Refs.~\cite{eckart08,arnold}, is the simple connection existing between the initial wave function in position representation for the atoms in an expanding BEC and the wave function in momentum representation at later times. This relationship is obtained by Fourier transforming the scaling solution~\eqref{eq:scaling}:   
\begin{widetext}
\begin{align}
\tilde{\psi} (\mathbf{p},t) &=
\big( \det \Lambda(t) \big)^{-\frac{1}{2}} \, e^{i \beta(t)} \int \frac{d^3x}{(2\pi \hbar)^{3/2}}\,
e^{-\frac{i}{\hbar} \mathbf{p}^\mathrm{T} \mathbf{x}} \,
e^{i\frac{m}{2\hbar}\, \boldsymbol{\xi}^\mathrm{T} A(t)\, \boldsymbol{\xi}} \,
\psi_\text{TF} (\boldsymbol{\xi}) \nonumber \\
&= \big( \det \Lambda(t) \big)^{\frac{1}{2}} \, e^{i \beta(t)} \int \frac{d^3\xi}{(2\pi \hbar)^{3/2}}\, 
e^{-\frac{i}{\hbar} \mathbf{p}^\mathrm{T} \Lambda(t)\, \boldsymbol{\xi}} \,
e^{i\frac{m}{2\hbar}\, \boldsymbol{\xi}^\mathrm{T} A(t)\, \boldsymbol{\xi}} \,
\psi_\text{TF} (\boldsymbol{\xi})
\label{D1},
\end{align}
\end{widetext}
where in the second equality we simply performed a change of integration variables. One can then use the same procedure employed in the evaluation of Eq.~\eqref{C1} by rewriting Eq.~\eqref{D1} as
\begin{align}
\tilde{\psi} (\mathbf{p},t) &=
e^{-\frac{i}{2 m \hbar} \mathbf{p}^\mathrm{T} \Lambda\, A^{-1} \Lambda^\mathrm{T} \mathbf{p}}
\int \! \frac{d^3\xi}{(2\pi \hbar)^{3/2}}\ (\det \Lambda)^\frac{1}{2}\, e^{i \beta}\,
\psi_\text{TF} (\boldsymbol{\xi}) \nonumber \\
& \qquad \qquad \times e^{i\frac{m}{2\hbar}\, \big(\boldsymbol{\xi} - A^{-1} \Lambda^\mathrm{T} \mathbf{p}/m \big)^\text{T} A \, \big(\boldsymbol{\xi} - A^{-1} \Lambda^\mathrm{T} \mathbf{p}/m \big)}
\label{D2},
\end{align}
and using the saddle-point approximation to obtain
\begin{align}
\tilde{\psi} (\mathbf{p},t) & \approx \,
\frac{i^{3/2}(\det \Lambda)^{1/2}\, e^{i \beta}}{m^{3/2}\, (\det A)^{1/2}}\;
e^{-\frac{i}{2 m \hbar} \mathbf{p}^\mathrm{T} \Lambda\, A^{-1} \Lambda^\mathrm{T} \mathbf{p}}
\nonumber \\ 
& \qquad \qquad \qquad \qquad \quad
\times \psi_\text{TF} \big( A^{-1} \Lambda^\mathrm{T} \mathbf{p}/m \big)
\label{D3},
\end{align}
which is valid when the widths of the Gaussian in Eq.~\eqref{D2} are smaller than the characteristic scales $\Delta \boldsymbol{\xi}$ of $\psi_\text{TF} (\boldsymbol{\xi})$, i.e.\ when $\sqrt{\hbar/(m |A_{ii}|)} \ll \Delta \xi^i \sim R_\text{TF}^{(i)}$ for $i=1,2,3$ in the basis where the matrix $A_{ij}$ diagonalizes.

It is interesting to consider the linear regime for the scaling matrix, where $\Lambda (t) = C + B\, (t-t_0)$. Whenever the time-dependent Thomas-Fermi approximation is valid, this corresponds to the situation where almost all the nonlinear interaction energy has already been converted into kinetic energy. In this case the condition for the validity of the saddle-point approximation becomes $R_\text{TF}^{(i)} \gg \sqrt{\hbar/(m |B_{ii}| |\Lambda_{ii}|)}$.
[Here we considered the basis where the matrices $B$ and $\Lambda(t)$ both diagonalize, which is always possible for nonrotating traps and coincides with the basis where the matrix $\Omega^2(t)$ diagonalizes.]
Note, for comparison, that Heisenberg's uncertainty principle implies $R_\text{TF}^{(i)} |\Lambda_{ii}|\, m\, R_\text{TF}^{(i)} |B_{ii}| \gtrsim \hbar$, i.e.\ $R_\text{TF}^{(i)} \gtrsim \sqrt{\hbar/(m |B_{ii}| |\Lambda_{ii}|)}$.
\comment{In terms of the phase-space description of the condensate's mean field this amounts to having a sufficient tilt of the associated Wigner function, as can be seen from Eq.~\eqref{eq:wigner_BEC2}.}
The tilt develops when the nonlinear interaction energy is converted into kinetic energy and it is always large enough for a BEC in the Thomas-Fermi regime which is released nonadiabatically. On the other hand, when using delta-kick cooling or a similar technique to reduce the momentum spread by rotating the Wigner function in phase space, the tilt is reduced (or even entirely removed) and one gets close to the saturation of Heisenberg's uncertainty relation. If this is done efficiently enough, the condition for the validity of the saddle-point approximation will no longer hold and the accuracy of Eq.~\eqref{D3} will no longer be guaranteed.
In that case, in fact, even the scaling approximation will cease to be valid if the condensate is dilute enough at that time because the expansion will then be mainly driven by the gradient of the envelope rather than by the kinetic energy that results from the conversion of the nonlinear interaction energy.

\subsection{Late-time scaling approach \emph{versus} free evolution}
\label{sec:scaling_vs_free}

The result of the previous subsection can be combined with that of Appendix~\ref{sec:late-time} to analyze the late-time behavior of an expanding BEC. In order to do so, we consider an intermediate time $t_1$ at which the density of the condensate is already low enough that we can neglect the nonlinear interactions. Hence, at this point the matrix $\Lambda(t)$ has already entered the linear regime. One can use Eq.~\eqref{D3} to evaluate $\tilde{\psi} (\mathbf{p},t_1)$, and substituting the expression for $\Lambda(t)$ linear in time, one gets:
\begin{equation}
\tilde{\psi} (\mathbf{p},t_1) \approx \,
\frac{(\det B)^{-\frac{1}{2}}}{(-i\,  m)^{3/2}} \;
e^{i \beta}\,
e^{-\frac{i}{\hbar} \mathbf{p}^2 \Delta t_1 / 2m } \;
\psi_\text{TF} \big( B^{-1} \mathbf{p}/m \big)
\label{D4},
\end{equation}
with $\Delta t_1 = (t_1 - t_0)$. Furthermore, for times $t \geq t_1$  the atoms in the condensate behave as free particles (since the interactions can be neglected) and the method of Appendix~\ref{sec:late-time} can be employed.
Making use of Eq.~\eqref{C1} applied to this case and taking Eq.~\eqref{D4} for  the initial wave function at time $t_1$ in momentum representation, one gets
\begin{align}
\psi (\mathbf{x},t) &=
\int \frac{d^3p}{(2\pi \hbar)^{3/2}}\, e^{i \beta}\,
e^{\frac{i}{\hbar} \mathbf{p}^\mathrm{T} \mathbf{x}} \,
e^{-\frac{i}{\hbar} (\mathbf{p}^2 / 2m) (t-t_1)} \, \tilde{\psi} (\mathbf{p},t_1)
\nonumber\\
&= \frac{(\det B)^{-\frac{1}{2}}}{(-i\, m)^{3/2}}
\int \frac{d^3p}{(2\pi \hbar)^{3/2}}\, e^{i \beta}\,
e^{\frac{i}{\hbar} \mathbf{p}^\mathrm{T} \mathbf{x}} \,
e^{-\frac{i}{\hbar} \mathbf{p}^2\, \Delta t / 2m} \, \nonumber \\
& \qquad \qquad \qquad \qquad \qquad
\times \psi_\text{TF} \big( B^{-1} \mathbf{p}/m \big)
\label{D5},
\end{align}
where in the second equality we combined the two phase factors with exponent proportional to $\mathbf{p}^2$ and introduced $\Delta t = (t-t_1) + \Delta t_1 = (t-t_0)$. Next, rewriting Eq.~\eqref{D5} as done in Eq.~\eqref{C2} and using the same kind of saddle-point approximation that led to Eq.~\eqref{C3}, one finally obtains:
\begin{equation}
\psi (\mathbf{x},t) \approx \big(\!\det (B \, \Delta t) \big)^{-\frac{1}{2}} \, e^{i \beta}\,
e^{\frac{i}{\hbar} \frac{m}{2 \Delta t} \mathbf{x}^2}\,
\psi_\text{TF} \big( B^{-1} \mathbf{x}/\Delta t \big) 
\label{D6},
\end{equation}
which coincides with the scaling solution~\eqref{eq:scaling} when $C$ can be neglected compared to $B\, \Delta t$.
The saddle-point approximation is valid provided that the typical scales of variation $\Delta \mathbf{p}$ of the function $\tilde{\psi} (\mathbf{p},t_1)$ satisfy the condition $\sqrt{m\hbar/\Delta t} \ll \Delta p^i$, which implies
$\sqrt{\hbar/(m |B_{ii}|^2 \Delta t)} \ll |B_{ii}|^{-1} \Delta p^i/m \sim R_\text{TF}^{(i)}$ for $i=1,2,3$.
For sufficiently late times (i.e.\ sufficiently large $\Delta t$) this condition will be fulfilled as long as the condition justifying the derivation of Eq.~\eqref{D3} holds.

The result found in this subsection is particularly interesting because it means that the validity of the scaling approximation can be automatically extended to late times during the free expansion: if it is valid at some time $t_1$ at which the expansion dynamics has already entered the linear regime, it will be valid at arbitrarily late times.


\bibliographystyle{apsrev4-1}
\bibliography{literature}

\end{document}